\documentclass[prb,aps,twocolumn,groupaddress]{revtex4-1}
\usepackage{latexsym}
\usepackage{graphicx}
\usepackage{verbatim}
\usepackage{multirow}
\usepackage{amsmath}
\usepackage{mathrsfs}
\usepackage{float}
\usepackage{feynmf}
\usepackage[usenames, dvipsnames]{color}

\newcommand{\be}{\begin{equation}}
\newcommand{\ee}{\end{equation}}
\newcommand{\ba}{\begin{eqnarray}}
\newcommand{\ea}{\end{eqnarray}}
\newcommand{\baa}{\begin{eqnarray*}}
\newcommand{\eaa}{\end{eqnarray*}}

\def\be{\begin{equation}}
\def\ee{\end{equation}}
\def\bea{\begin{eqnarray}}
\def\eea{\end{eqnarray}}

\def\C60{A$_x$C$_{60}$}

\def\HgCu3{HgCa$_2$Cu$_3$O$_{8+y}$}
\def\HgCu4{HgBa$_2$Ca$_3$Cu$_4$O$_{10+y}$}
\def\TlCu{Tl$_2$Ba$_2$CuO$_{6+\delta}$}
\def\TlCu3{Tl$_2$Ba$_2$Ca$_2$Cu$_3$O$_{10+y}$}
\def\TlCu4{Tl$_2$Ba$_2$Ca$_3$Cu$_4$O$_{12+y}$}

\def\BiCu3{Bi$_2$Sr$_2$Ca$_{2}$Cu$_3$O$_y$}

\def\8LSCO{La$_{1.88}$Sr$_{.12}$CuO$_4$}
\def\110LNSCO{La$_{1.5}$Nd$_{0.4}$Sr$_{0.1}$CuO$_{4}$}
\def\stage4LCO{La$_{2}$CuO$_{4+\delta}$}
\def\Y248{YBa$_2$Cu$_4$O$_8$}

\def\NbSe2{NbSe$_2$}
\def\TaSe2{TaSe$_2$}
\def\TiSe2{TiSe$_2$}

\begin{document}
\title{Low energy inelastic response in the superconducting phases of $\mathbf{PrOs_4Sb_{12}}$}
 \author{Chandan Setty}
\affiliation{Department of Physics and Institute for Condensed Matter Theory, University of Illinois 1110 W. Green Street, Urbana, IL 61801, USA}
 \author{Yuxuan Wang}
\affiliation{Department of Physics and Institute for Condensed Matter Theory, University of Illinois 1110 W. Green Street, Urbana, IL 61801, USA}
\author{Philip W. Phillips}
\affiliation{Department of Physics and Institute for Condensed Matter Theory, University of Illinois 1110 W. Green Street, Urbana, IL 61801, USA}
\begin{abstract}
 Recent AC susceptibility and polar Kerr effect measurements in the skutterudite superconductor $\mathrm{PrOs_{4}Sb_{12}}$ (POS) [Levenson-Falk et. al arXiv:1609.07535] uncovered the nature of the superconducting double transition from a high temperature, high field, time reversal symmetric phase (or the A phase) to a low temperature, low field, time reversal symmetry broken phase (or the B phase). Starting from a microscopic model, we derive a Ginzburg-Landau expansion relevant to POS that describes this entrance into the time reversal symmetry broken phase along the temperature axis. We also provide a study of the low energy inelastic (Raman) response in both the A and B phases of POS, and seek additional signatures which could help reveal the exact form of the gap functions previously proposed in these phases. By appropriately manipulating the incoming and scattered light geometries, along with additional subtraction procedures and suitable assumptions, we show that one can access the various irreducible representations contained in the point group describing POS. We demonstrate how to use this technique on example order parameters proposed in POS. Depending on whether there exist nodes along the $c-$ axis, we find additional low energy spectral weight within the superconducting gap in the $E_g$ geometry, a feature that could pin point the location of nodes on the Fermi surface.
\end{abstract}
\maketitle
\section{Introduction} Ever since its original discovery early last decade \cite{Maple2002,Lynn2002}, the heavy fermion skutterudite superconductor  $\mathrm{PrOs_{4}Sb_{12}}$ has met every measure used to quantify unconventionality in superconductors \cite{Yanagisawa2006-Review,Sato2007-Review}. Two distinct specific heat jumps have been observed \cite{Maple2003-SpecificHeat} at temperatures close to $T_c \sim 1.85 K$ and $T_ c \sim 1.7 K$ indicating two different transitions to the superconducting state in the absence of a magnetic field. This conclusion has been backed by additional thermal expansion \cite{Maple2003-ThermalExpansion} and transport \cite{Maki2003} measurements which report identical jump-like features supporting a double transition to the superconducting state. The power-law behavior of the Nuclear Magnetic Resonance (NMR) spin-lattice relaxation rate \cite{Sato2005,Sato2011}, Nuclear Quadrupole Resonance (NQR) \cite{katayama2007evidence}, penetration depth \cite{chia2003probing} and specific heat \cite{Maple2003-SpecificHeat,Maple2002,Lynn2002} seem to suggest a point-nodal superconducting gap-like behavior, although other thermal conductivity, muon spin resonance ($\mu$SR) and NQR  studies seem to support a fully gapped Fermi surface~\cite{seyfarth2006superconducting, maclaughlin2002muon,Sato2003}. The presence of a non-zero Kerr angle \cite{Sato2003-TRSB,Kapitulnik2016} points toward a chiral time reversal symmetry broken (TRSB) superconducting ground state, with a more recent measurement \cite{Kapitulnik2016} reporting that the TRSB state is specific to the low temperature, low magnetic field superconducting phase (B phase) and absent in the high temperature, high magnetic field phase (A phase). Additionally, Knight shift measurements \cite{Sato2007-SpinTriplet} have shown signs of spin triplet pairing and, along with TRSB, has motivated a recent proposal \cite{Liang2016} suggesting that POS could prove to be a promising candidate for a chiral superconductor hosting three-dimensional Majorana fermions. In the normal state of POS, low temperature specific heat measurements \cite{Sato2002-FieldInduced} in the presence of a magnetic field have identified a broad magnetic field induced ordered phase characterized by an enhancement of the 4$f$ magnetic moment.    \\
\newline
A fundamental question concerning the character of the superconducting ground state is the pairing mechanism and symmetry of the order parameter defined on the Fermi surface. Several theoretical efforts have already been put forth to classify and probe the pairing mechanism and symmetry in POS \cite{Curnoe2004,Matsuda2003,Matsuda2004,Goryo2003,Harima2003-Quadrupolar, Machida2003}. The structure of the gap functions in the spin singlet pairing channel have been considered \cite{Goryo2003} using a Ginzburg-Landau (GL) phenomenological approach in both the A and B phases. In the absence of any spin orbit coupling, it was concluded that in the A phase a strongly anisotropic $s$-wave is favored; using the basis functions for the irreducible representations of the $T_h$ point group (see the Character table in Table \ref{Character}), a single order parameter component of the forms $\Delta(\vec k) = s + k_x^4 + k_y^4 + k_z^4$ or $\Delta(\vec k) = s + k_x^2  k_y^2 +  k_y^2  k_z^2 + k_z^2 k_x^2 $ with an $A_g$ symmetry were argued to be competent ($s$ is a scalar number). In the B phase, a TRSB $s+ id$ state with a gap structure of the form $\Delta(\vec k) = ( s + k_x^2  k_y^2 +  k_y^2  k_z^2 + k_z^2 k_x^2 ) $  +  $i (k_x^2 - k_y^2)$, with combinations of the one dimensional $A_g$ and two dimensional $E_g$ representations was argued to be the ground state. In the presence of spin orbit coupling, an $f$-wave pairing state with point nodes along all the three axes in momentum space was suggested in \cite{Machida2003}, an novel $s+g-$ wave pairing was found in ref \cite{Matsuda2003}, and, using a microscopic approach with quadrupole fluctuation mediated pairing, Miyake and coworkers \cite{Harima2003-Quadrupolar} found a chiral $p_x + i p_y$ pairing order parameter. Several other order parameters were proposed \cite{Curnoe2004} based on a rigorous symmetry analysis in which the authors, following the work of Volovik and Gor'kov \cite{Gorkov1985}, minimized a Landau free energy functional that was invariant under the point group, gauge and time reversal symmetries, and studied the ensuing gap nodal structures.  \\
\newline
Electronic Raman scattering in superconductors \cite{Dierker1984,Hackl2007,Einzel1995,Neumeier1994,Zawadowski1996} has proved to be an indispensable tool for determining the structure and symmetry of the gap in unconventional, anisotropic superconductors. Light is incident with a certain initial electric polarization and energy ($\hat{e}^i, \omega_I$) and scattered with a final polarization and energy ($\hat{e}^s, \omega_S$). The energy difference corresponds to the energy needed to break a Cooper pair or to excite a low energy in-gap mode.  The Raman scattering vertex, in general, can have contributions from all the irreducible representations (IRRP), $\Gamma_i$, of the point group of the crystal; however, a proper choice of the polarization geometries of the incoming and outgoing light beams can single out contributions from a single IRRP to the total scattering cross section. As a result of this selective Brillouin zone averaging, the quasiparticle energy gaps are probed only in certain directions in the Brillouin zone. Thus this technique
is sensitive to the position of gap nodes, and therefore provides
indirect information about the symmetry of the gap function and its anisotropies. If there are several candidate pairing symmetries with different location of nodes on the Fermi surface, this method can (at least) help narrow down the various possible candidates.\\
\newline
In this work, starting from a microscopic model, we derive a GL expansion relevant to POS that shows its entry into the TRSB phase at low temperatures. We also estimate the GL coefficients using their relationships to the various microscopic parameters. As an experimental means to probe the proposed pairing symmetries, we examine the low energy inelastic (Raman) response in both the A and B phases of POS; in particular, we seek to qualitatively understand its dispersive features for low energy and zero momentum transfer. By appropriate manipulation of the incoming and scattered light geometries, as well as other subtraction procedures routinely applied \cite{Einzel1995} in this method, we demonstrate that one can probe the various irreducible representations contained in the $T_h$ point group describing POS. For the purposes of illustration, we stick to the spin singlet, even parity forms of the gap functions proposed by Goryo \cite{Goryo2003} in the A and B phases, as was discussed in the preceeding paragraphs.  Depending on the existence of nodes along the $c-$ axis, we find an enhancement (nodal) or suppression (gapped) of the low-energy spectral weight in the $E_g(1)$ and $E_g(2)$ geometries (here (1), (2) etc. denote the individual elements of the respective multi-component IRRP-). This constitutes a defining property in electronic inelastic light scattering that could help pin-point the exact location of nodes on the fermi surface and, thereby, narrow down the candidate pairing symmetries proposed for POS. \\
\newline
\begin{figure}[h!]
\includegraphics[width=1.7in,height=1.6in]{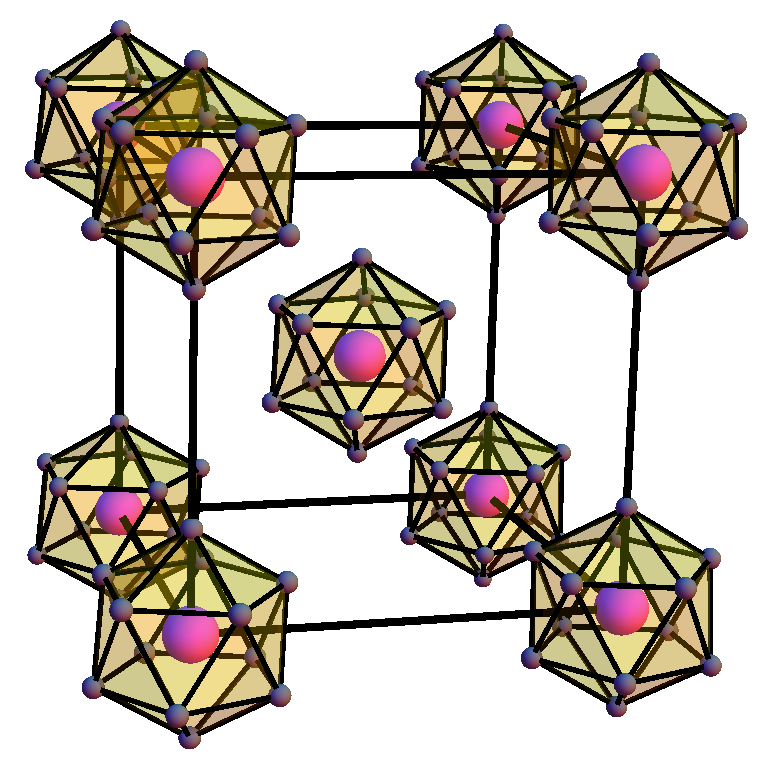}\hfill%
\includegraphics[width=1.7in,height=1.5in]{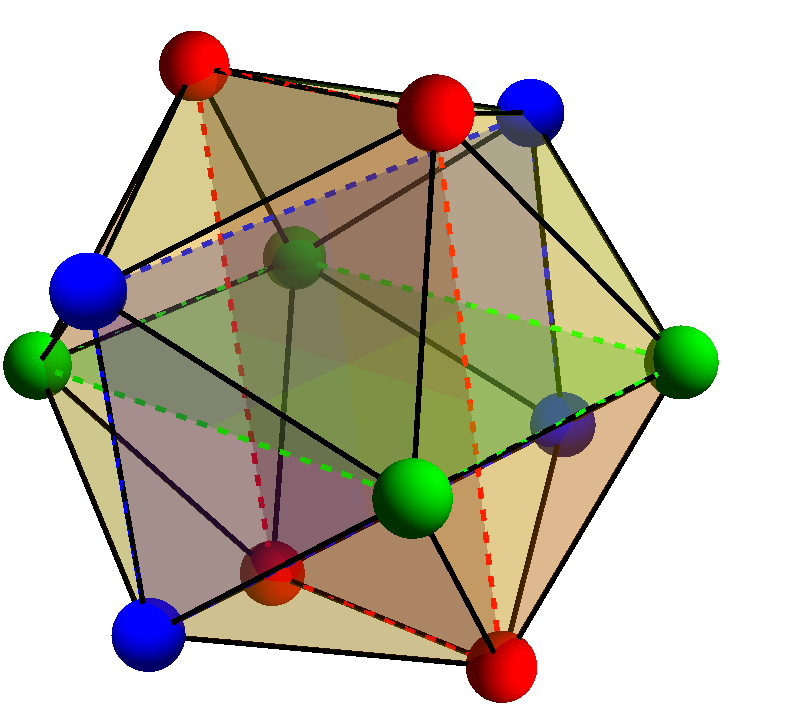}%
\caption{(Left) Lattice structure of the skutterudite $PrOs_4Sb_{12}$ which belongs to the $Im\bar 3$ (Tetrahedral, $T_h$) space group (only the Pr (magenta) and Sb (gray) atoms are shown). The Pr atoms form a body centered cubic lattice while the Sb atoms form an icosahedral cage surrounding each Pr atom. (Right) Closer image of the icosahedral cage (Black solid lines) formed by twelve Sb atoms. This can be thought of as three rectangular planes (red, green and blue dashed lines) orthogonal to each other and intersecting at a common origin. }\label{POSLattice}
\end{figure}
\section{Lattice and Fermi surface} 
\begin{figure}[h!]
\includegraphics[width=1.7in,height=2in]{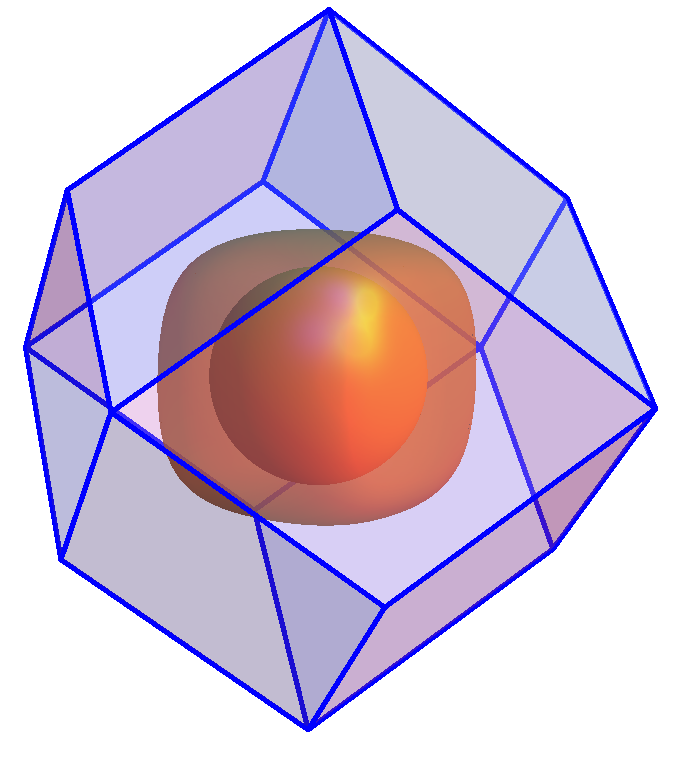}\hfill%
\includegraphics[width=1.7in,height=2in]{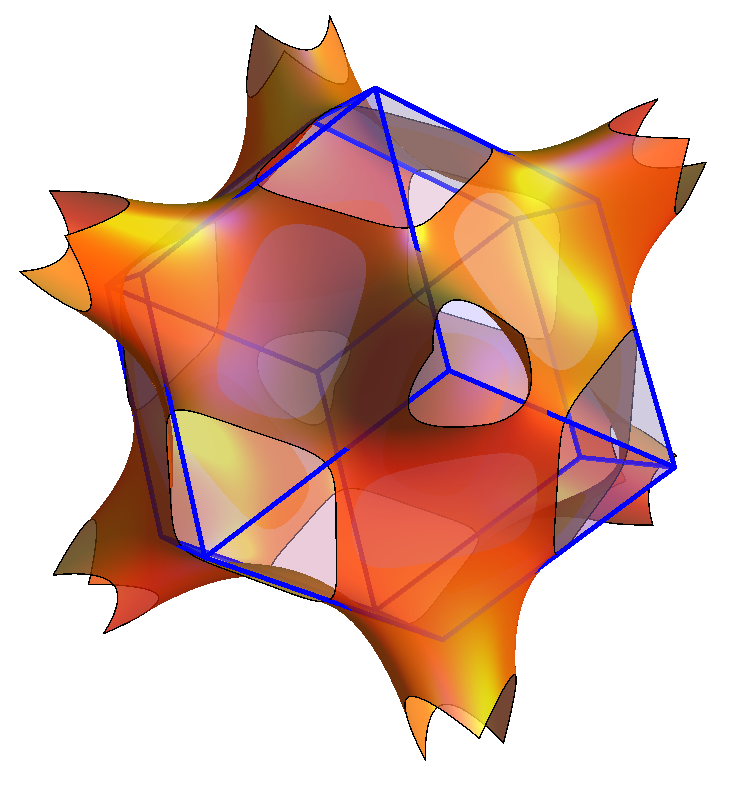}%
\caption{Fermi surfaces obtained from the basis functions appearing in the character table for the $T_h$ point group (see Table \ref{Character}). The enclosed volume denotes the Brillouin zone for the body centered cubic lattice (a dodecahedron). (Left) Two Fermi hole pockets centered around the $\Gamma$ point: an inner spherical pocket ($\epsilon_1(\vec k)$) and an outer rounded cubic pocket ($\epsilon_2(\vec k)$). (Right) A multiply connected pocket with parts of its surface intersecting the $N$ point (mid point of each face of the dodecahedron) obtained from $\epsilon_3(\vec k)$.   }\label{FermiSurface}
\end{figure}

The lattice structure of POS is shown in Fig \ref{POSLattice}(left). The relevant space group of interest is the $Im\bar{3}$ (No. 204)  with a tetrahedral $T_h$ point group. The lattice structure consists of Pr atoms (shown in magenta) forming a body centered cubic lattice and the Sb atoms (shown in gray balls) forming icosahedral cages (yellow structures) encapsulating the Pr atoms. The Os atoms (not shown for purposes of clarity) form a cube and are intercalated within the body centered cubic structure. The icosahedral cage is shown in more detail in Fig \ref{POSLattice} (right). Each cage consists of three rectangular planes (shown in red, green and blue) of Sb atoms lying perpendicular to each other, along the three coordinate planes, with a common intersection point at the origin. The symmetry group describing of the Pr atoms, by themselves, is the cubic $O_h$ group, while that of the Sb atoms is the tetrahedral $T_h$ group. Since the cubic $O_h$ is a larger symmetry compared to the tetrahedral $T_h$ (all symmetry elements in $T_h$ are also present in $O_h$), the symmetry of the entire lattice is determined by the smaller $T_h$ point group.
 One should, therefore, use the irreducible representations of $T_h$ (for example, see refs \cite{Dresselhaus2007,Cardona2005}) to construct invariants to describe POS (this is made possible because the symmetry axes belonging to the Sb icosahedral cages coincide with that of the Pr body centered cubic lattice ).  \\
\newline
Several quantum oscillation measurements \cite{Harima2002, Onuki2009, Kobayashi2008} have been performed to map out the Fermi surface structure of POS. The Brillouin zone is a rhombicdodecahedron (just as in the case of a body centered cubic lattice), with its origin at the $\Gamma$ point and the center of the rhombic face labelled as the $N$ point (for more details about the other high symmetry points and axes, the reader can refer to \cite{Cracknell2010}). All quantum oscillation measurements observe two hole pockets centered around the $\Gamma$ point $-$ an inner spherical pocket and an outer rounded cubic pocket. There is another larger multiply connected pocket, centered again around the $\Gamma$ point, and touching the face of the boundary rhombus at the $N$ point. The Brillouin zone and the Fermi surface of POS is shown in fig \ref{FermiSurface}.
\begin{table}
\begin{tabular}{l*{6}{c}r}
$T_h$  & E & $3 C_2$ & $8 C_3$ & $i$ & $3 i C_2$  & $8 i C_3$ & Basis \\
\hline
$A_g$ & 1 & 1 & 1 & 1 & 1 & 1 & $r_s^2 + z^2$  \\
$E_g$  & 2 & 2 & -1 & 2 &  2 &-1 &  ($ r^2_a$,$2 z^2 - r_{s}^2$)  \\
$T_g$  & 3 & -1 & 0 & 3 &  -1 & 0 &  $(xz, yz, xy)$  \\
$A_u$  & 1 & 1 & 1 & -1&  -1 & -1 &  - \\
$E_u$  & 2 & 2 & -1 & -2 &  -2 & 1 &  -  \\
$T_u$  & 3 & -1 & 0 & -3 &  1 & 0&  $(x, y, z)$ \\
\end{tabular}
\caption{Character table for the $T_h$ point group. In the basis function column we have defined $r_s^2 = x^2 + y^2$ and $r_a^2 = x^2 - y^2$.}\label{Character}
\end{table}
 Currently, we do not know of any exhaustive study attempting to write down a simple tight-binding model that captures all the essential kinematics such as the band structure, Fermi surface and orbital character of POS. However, to study the electronic Raman response of a material, we only need the underlying symmetry properties of POS. Hence, for the purposes of this paper, we will only be interested in the symmetry characters of the individual bands and not other microscopic details. To achieve this, and at the same time maintain analytical and numerical tractability of the problem, we will work in the band basis and ignore any interband effects. This is a reasonable assumption as the low energy interband response can be neglected for widely spaced bands.  To construct these dispersions, we will rely on the basis functions appearing in the character table presented in Table \ref{Character}. For the three Fermi surfaces shown in Fig \ref{FermiSurface}, we simply construct invariants (i.e functions that transform as a singlet $A_g$) with respect to symmetry operations of the $T_h$ point group. We have chosen the invariants such that the Fermi surfaces match the experimentally measured contours in refs \cite{Harima2002, Onuki2009, Kobayashi2008} through quantum oscillations. These invariants are given as
\begin{eqnarray} \nonumber
\epsilon_1(\vec k) &=& \epsilon_1 + A_{g}(\vec k) - \mu\\ \nonumber
\epsilon_2(\vec k) &=& \epsilon_2 + A_{g}(\vec k) - 2 \vec{E}_g(\vec k)\cdot\vec{E}_g(\vec k) - \mu \\
\epsilon_3(\vec k) &=&  \epsilon_3 - \vec{T}_{u}(\vec k)\cdot \vec{T}_{u}(\vec k) - \mu
\label{Dispersions}
\end{eqnarray}
 where $(\epsilon_1, \epsilon_2, \epsilon_3) = (-6.88,- 6.7, 3) eV$,  $\mu = -1 eV$,  and the basis functions are given by 
\begin{eqnarray}\nonumber
A_g(\vec k) &=&  2 (c_x +  c_y +  c_z)\\ \nonumber
\vec{E}_g(\vec k) &=& 2\left( c_x -  c_y, 1- c_z - \frac{1}{2}(2-  c_x -  c_y) \right)\\
\vec{T}_u(\vec k) &=& 4\left( s_{\frac{x}{2}}(c_y -  c_z), s_{\frac{y}{2}}(c_z -  c_x), s_{\frac{z}{2}}(c_x - c_y) \right). \label{Irreps}
\end{eqnarray}
Here, we have defined $c_x = \cos(k_x)$, $s_{\frac{x}{2}} = \sin(k_x/2)$ and so on. It is easy to check that Eqs~\eqref{Irreps} transform according to their respective irreducible representations (albeit in a lattice version) as specified in the character table in Table \ref{Character}. It should also be noted that in the last equation for $\vec T_{u}(\vec k)$, we have used higher order basis functions that are not present in Table \ref{Character}. This state of affairs arises because writing the dot product invariants with the lowest order basis functions of the $T_u$ representation yields a function equivalent to $\epsilon_1(\vec k)$. The three dispersions in Eqs. \ref{Dispersions} provide the Fermi surface contours in Fig \ref{FermiSurface}.
\\ \newline
\section{Time reversal symmetry breaking}
 In the following analysis, we derive a generalized GL expansion to understand time reversal symmetry breaking in POS. 
 On the theoretical side\cite{Shiina2004-Multipolar,Aoki2004-FieldInducedAFQ,Koga2004-AFQandNeutron}, the effect of magnetic and quadrupole moments on the phase diagram was outlined. Motivated by the evidence of such magnetic moments  \cite{Sugawara2003-Neutron, KuwaharaSato2004, Lynn2002,Maple2002,Sato2003-Magnetization,Sato2002-FieldInduced}, magnetic exchange correlations, quadrupole moments \cite{Sugawara2003-Neutron,Sato2003-Magnetization,Sato2002-FieldInduced,RotunduSato2004-PRL} and quadrupole exchange correlations, and taking into account the crystal symmetries appropriate for quartic interactions generated by such correlations, we start with the following form of the action written in momentum space
\begin{widetext}
\begin{eqnarray}
S(\bar c, c) &=& \beta \sum_{ \substack{\alpha\beta \\ \textbf k \sigma }} \bar c_{\alpha \sigma}(\textbf k)\left(- ik_n\delta_{\alpha\beta} - \epsilon_{\alpha \beta}(\vec k)\right) c_{\beta \sigma}(\textbf k) - \beta g \sum_{\substack{\textbf k \textbf k' \alpha}} \bar c_{\alpha \uparrow}(\textbf k)  D(\vec k)\bar c_{\alpha \downarrow}(-\textbf k) c_{\alpha \downarrow}(-\textbf k')  D(\vec k')   c_{\alpha \uparrow}(\textbf k').  
\end{eqnarray}
\end{widetext}
Here  $\bar c_{\alpha \sigma}(\textbf k)$ and  $c_{\beta \sigma}(\textbf k)$ are the creation and annihilation Grassmann numbers that follow Grassmannian algebra for electrons with Bloch momentum $\vec k$ and Matsubara frequency $k_n$ (collectively denoted by $\textbf{k}$), spin $\sigma$ and orbital indices $\alpha, \beta$, $\epsilon_{\alpha \beta}(\vec k )$ are the orbital matrix elements, $g$ is the interaction strength which is taken to be a constant, $\beta$ is the inverse temperature, and $D(\vec k) \equiv \phi_s(\vec k) + \phi_d(\vec k)$ is the total form factor which is a mixture of the $s-$ wave and $d-$wave form factors. Here we have defined $\phi_s(\vec k) = \Delta_s \Phi_s(\vec k) $ and $\phi_d(\vec k) = \Delta_d \Phi_d(\vec k)$, where $\Delta_s$ and $\Delta_d$ are complex scalar amplitudes, along with $\Phi_s(\vec k) = (  k_x^2  k_y^2 +  k_y^2  k_z^2 + k_z^2 k_x^2 ) $ and $ \Phi_d(\vec k) =  (k_x^2 - k_ y^2)$.  The spin structure in the interaction term is chosen since we anticipate a BCS-like singlet superconducting order parameter. At this point, from an experimental perspective, it is not fully clear if the gaps on the different orbitals need to be distinct or not; hence from now on, for simplicity, we will choose the total form factor $D(\vec k)$ to be independent of the orbital index $\alpha$. Such an assumption will allow us to work in a basis where the kinetic energy and the order parameter matrices appearing in the Hamiltonian are both diagonal because they commute and hence, will simplify the problem to a collection of independent bands with the same superconducting gap on each band. Note that this is not always true in generic multiband superconductors (for example, the Iron- based superconductors) where the gap and kinetic energy matrices do not commute. Changing to the band basis with energy eigenvalues $\epsilon_{\alpha}(\vec k)$ by using the unitary transformation, $c_{\alpha\sigma}(\textbf k) = U_{\alpha\beta}(\vec k)\psi_{\beta \sigma}(\textbf k)$, performing a Hubbard-Stratonovich transformation to decouple the quartic terms in the action, and integrating out the fermionic degrees of freedom results in a free energy in the superconducting state of the form (see Appendix and Ref. \cite{Wang2017})
\begin{widetext}
\begin{eqnarray}\nonumber
\mathscr{F}_s &=& \alpha_s |\Delta_s|^2 + \alpha_d | \Delta_d |^2 + \beta_s  |\Delta_s |^4 +\beta_d | \Delta_d |^4 + 4 \beta_1 | \Delta_s |^2 | \Delta_d |^2  \\ \nonumber
&&+ \alpha' (\Delta_s^* \Delta_d + \Delta_s \Delta_d^*) + \beta_2 \left(\Delta_s^2 \Delta_d^{* 2} + \Delta_s^{*2} \Delta_d^{2} \right)  + 2 \sum_{\nu=\pm}g_{\nu} (| \Delta_s |^2 + \nu | \Delta_d |^2)  ( \Delta_s \Delta_d^* + 2 \Delta_s^* \Delta_d),
\label{FreeEnergy}
\end{eqnarray}
\end{widetext}
\begin{figure*}[t]
\includegraphics[width=1.75in,height=1.7in]{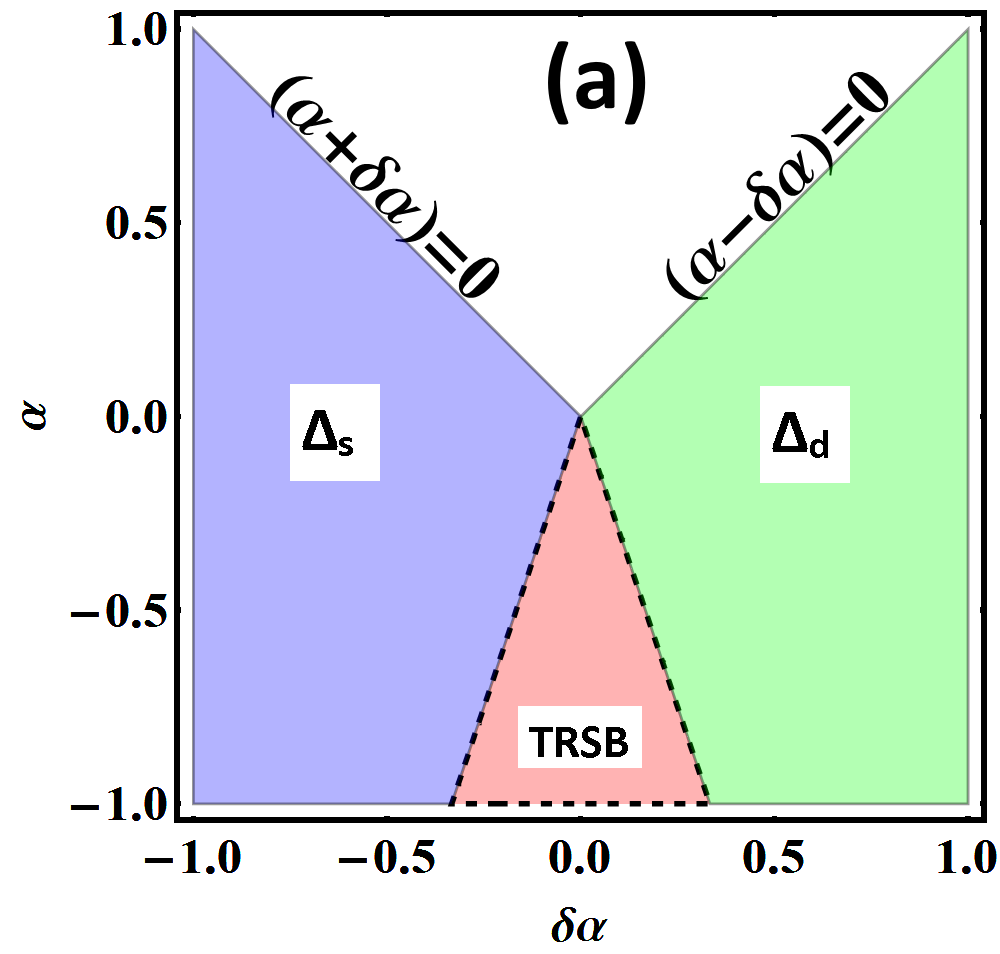}\hfill%
\includegraphics[width=1.75in,height=1.7in]{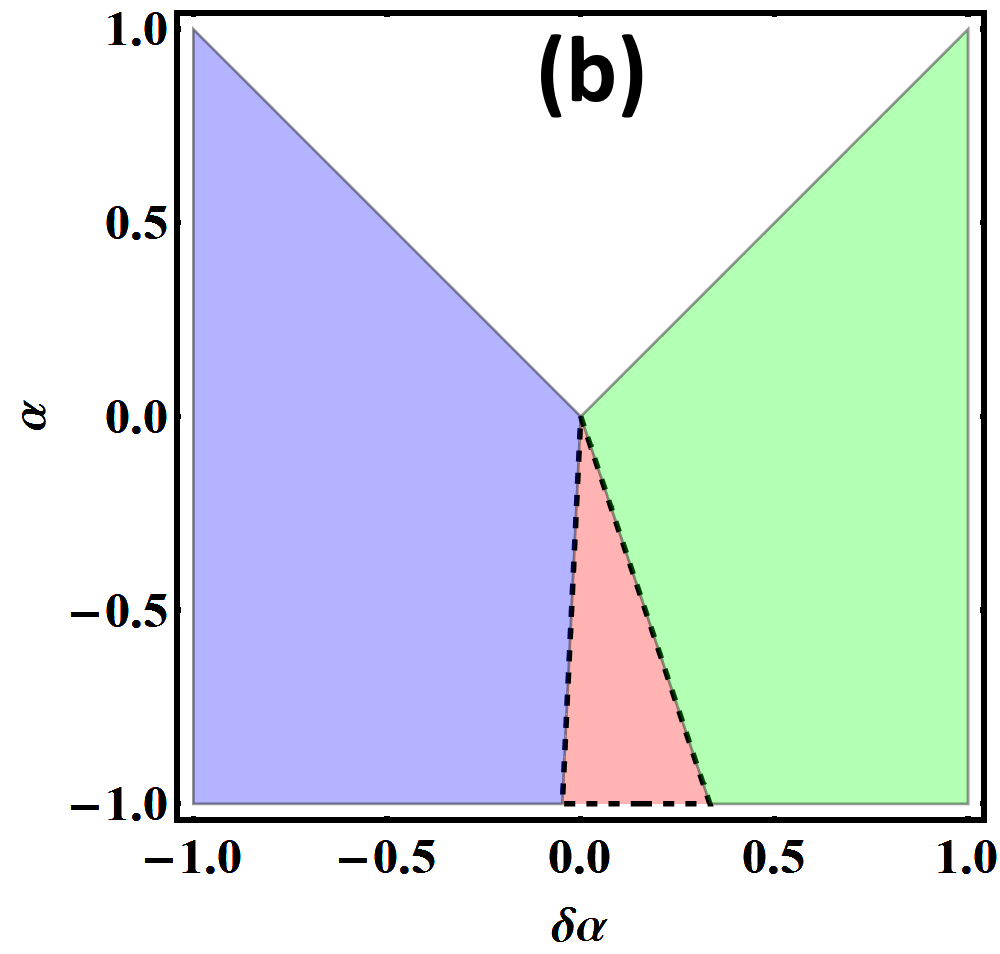}\hfill %
\includegraphics[width=1.75in,height=1.7in]{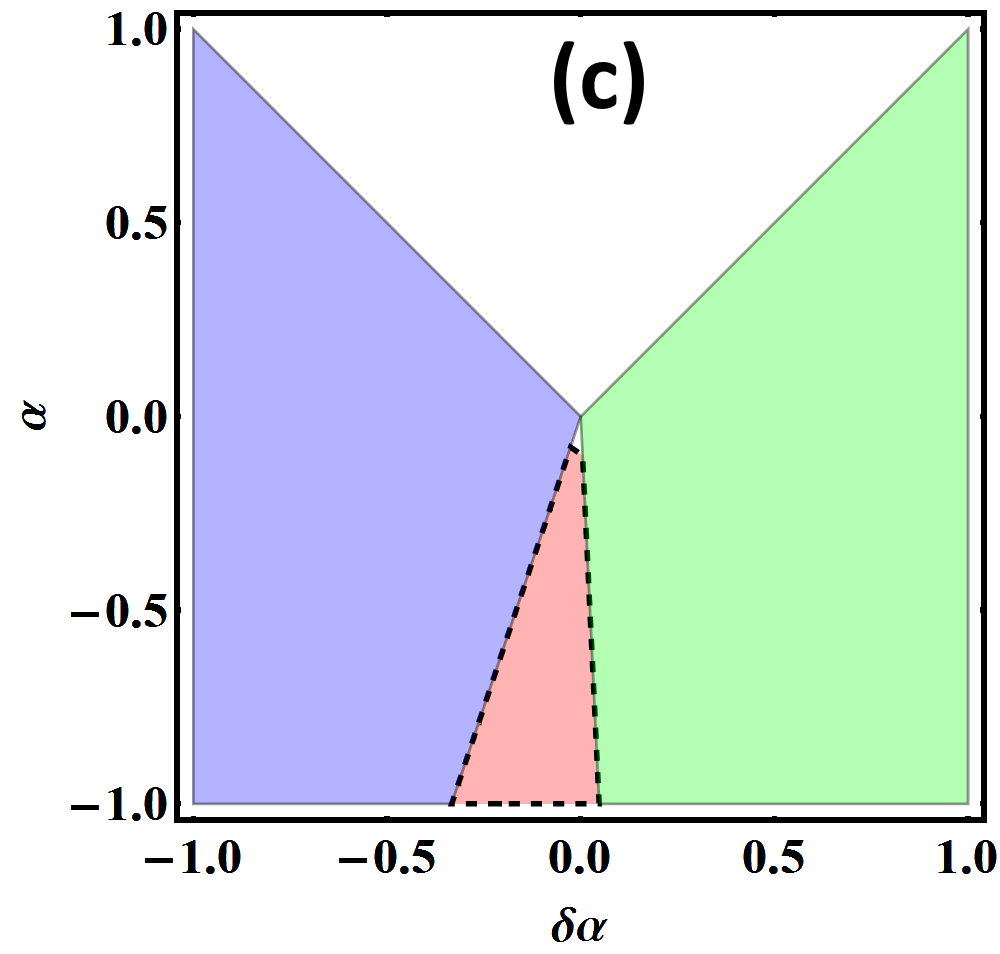}\hfill%
\includegraphics[width=1.75in,height=1.7in]{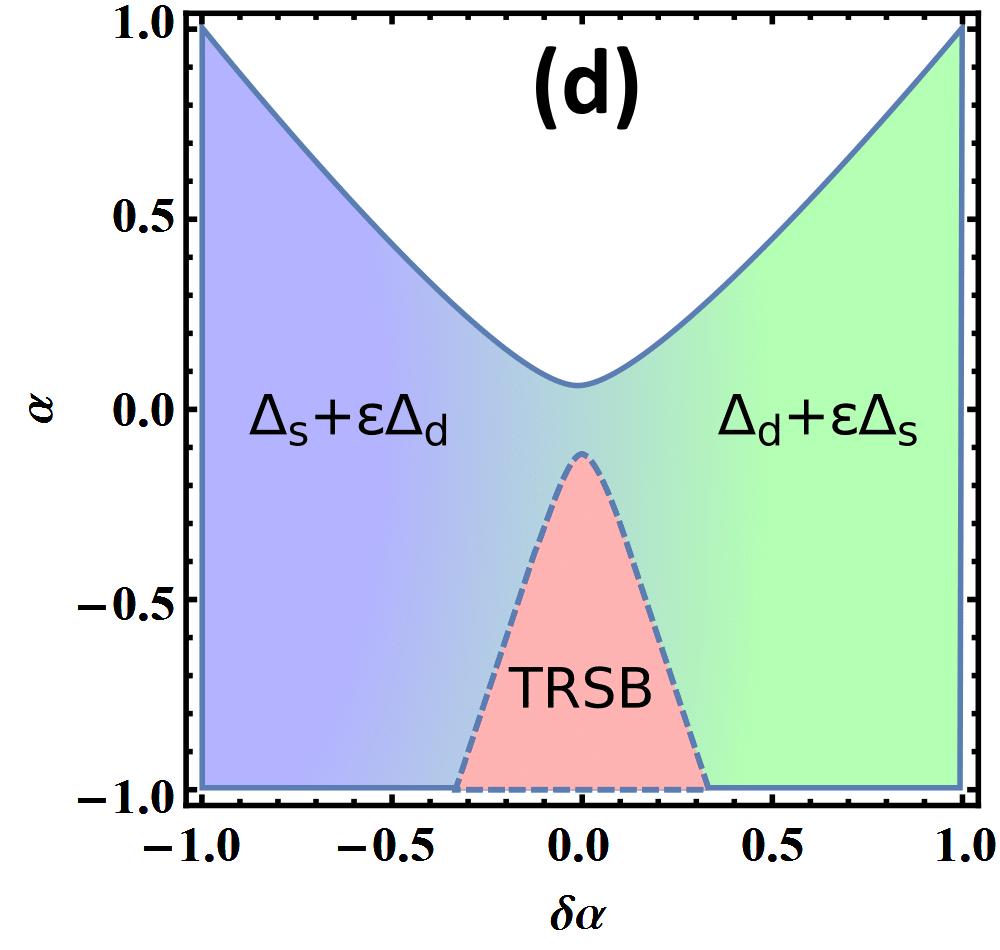}\hfill%
\caption{Phase diagram of the competing $s-$ and $d-$wave phases for the free energy appearing in Eq.\ref{FreeEnergy}. Panels (a-c) correspond to the case when the linear terms  proportional to $ (\Delta_s^* \Delta_d + \Delta_s \Delta_d^*)$ are ignored: (panel a) when $\beta_s = \beta_d$, (panel b) when $\beta_s>\beta_d$ and (panel c) when $\beta_s<\beta_d$. The case of POS corresponds to $\beta_d> \beta_s$. The labels in panels (b) and (c) follow along the lines of panel (a). The TRSB phase is bounded by the inequalities $\frac{\alpha(\beta_1 - \beta_d) - \delta \alpha (\beta_1 + \beta_d)}{2 (\beta_s\beta_d - \beta_1^2)}>0$ and $\frac{\alpha(\beta_s - \beta_1) - \delta \alpha(\beta_s + \beta_1)}{2(\beta_1^2 -\beta_s\beta_d)}>0$. Panel (d) corresponds to the case where the linear terms are included for $\beta_s = \beta_d$.  $\epsilon$ is a number much smaller than unity.}
\label{TRSBPhaseDiagram}
\end{figure*}
with the detailed form of the coefficients given in the Appendix. For the case of the $s-$ and $d-$wave symmetries chosen in our paper for POS,  $\alpha'$ and $g_{\pm}$ are small compared to the rest of the coefficients due to negligible overlap between odd powers of the $s-$ and $d-$ wave form factors, and hence can be neglected in the lowest-order approximation. We will later include these terms to see their effects on the phase diagram.  If we define the relative phase between $\Delta_s$ and $\Delta_d$ to be $\eta$, minimization of the free energy with respect to $\eta$ fixes $\eta=\pi/2$. Minimizing with respect to $|\Delta_s |$ and $| \Delta_d |$ gives us four sets of solutions of the form, $\left(|\Delta_s|, |\Delta_d|\right) = (0,0)$,  $\left(|\Delta_s|, |\Delta_d|\right) = (0,\sqrt{\frac{-(\alpha - \delta \alpha)}{2 \beta_d}})$, $\left(| \Delta_s|, | \Delta_d |\right)= (\sqrt{\frac{-(\alpha + \delta \alpha)}{2 \beta_s}},0)$, $\left(|\Delta_s|, |\Delta_d|\right) = (D_s,D_d)$, where
\begin{eqnarray}
 D_s &=& \sqrt{\frac{\alpha(\beta_1 - \beta_d) - \delta \alpha (\beta_1 + \beta_d)}{2 (\beta_s\beta_d - \beta_1^2)}}\\
D_d &=& \sqrt{\frac{\alpha(\beta_s - \beta_1) - \delta \alpha(\beta_s + \beta_1)}{2(\beta_1^2 -\beta_s\beta_d)}}.
\end{eqnarray}
 Here we have defined $\alpha = (\alpha_s + \alpha_d)/2 $ and $\delta \alpha = (\alpha_s - \alpha_d)/2$. The last pair of solutions where the relative phase between $\Delta_s$ and $\Delta_d$ is fixed by the $\beta_2$ term at $\pm\pi/2$ gives rise to the TRSB phase. Such a phase has the lowest free energy but is obviously a stable solution only when $(|\Delta_s |, | \Delta_d |)$ are both real. Hence, the conditions $\frac{\alpha(\beta_1 - \beta_d) - \delta \alpha (\beta_1 + \beta_d)}{2 (\beta_s\beta_d - \beta_1^2)}>0$ and $\frac{\alpha(\beta_s - \beta_1) - \delta \alpha(\beta_s + \beta_1)}{2(\beta_1^2 -\beta_s\beta_d)}>0$ define the TRSB phase. The $s-$wave and $d-$wave pairings are stable when $-(\alpha+\delta \alpha)$ and $-(\alpha-\delta \alpha)$ are positive respectively. These phases are shown in  Fig~\ref{TRSBPhaseDiagram}(a-c) for $\beta_s = \beta_d$ (Fig~\ref{TRSBPhaseDiagram}(a)), $\beta_s > \beta_d$ (Fig~\ref{TRSBPhaseDiagram}(b)) and $\beta_s< \beta_d$ (Fig~\ref{TRSBPhaseDiagram}(c)). Given the candidate pairing symmetries we have chosen for POS, viz. $\Phi_s(\vec k) $ and $ \Phi_d(\vec k) $, we find that POS belongs to the case where $\beta_s<\beta_d$. \\ \newline 
The presence of linear terms with the coefficients $\alpha'$ and $g_{\pm}$ reflects the fact that the $s$-wave and $d$-wave order parameters are \emph{not} distinguished by symmetry. Then a transition between $s$-wave and $d$-wave states does not require further breaking any symmetries -- in fact, the $s$ and $d$-wave components are always mixed even in the time-reversal invariant phase. Thus when $\Delta_s\sim \Delta_d$, the two order parameters coexist in two competing ways, either preserving or breaking time-reversal symmetry. Since the linear coupling terms are already present at the quadratic level, while the terms responsible for time-reversal symmetry breaking enter at the quartic level, the mixing of the two orders breaks time-reversal only at low-temperatures when order parameters have larger expectation values. Treating the $\alpha'$ and $g_{\pm}$ terms perturbatively, we obtain a schematic phase diagram shown in Fig~\ref{TRSBPhaseDiagram}(d). A more detailed calculation for obtaining this phase diagram is found in Ref~\cite{Chubukov2013}.\\ \newline
\section{Inelastic (Raman) scattering} The scattered light intensity is written in terms of the differential scattering cross section \cite{Dierker1984,Hackl2007,Einzel1995,Neumeier1994,Zawadowski1996} as
\begin{eqnarray}
\frac{\partial^2 \sigma}{\partial \omega \partial \Omega}&=&\frac{\omega_S}{\omega_I}r_0^2 S_{\gamma \gamma}(\vec{q},\omega)\\
S_{\gamma \gamma}(\vec{q},\omega) &=& -\frac{1}{\pi}[1+ n_B(\omega)] Im \chi_{\gamma \gamma}(\vec{q},\omega), \label{eq:CrossSection}
\end{eqnarray}
where $n_B(\omega)$ is the Bose-Einstein distribution function, $\omega_S$ and $\omega_I$ are the frequencies of the scattered and incident light respectively and $r_0 = \frac{e^2}{m c^2}$. The imaginary part of the inelastic response $ Im \chi_{\gamma \gamma}(\vec{q},\omega)$ (the subscript $\gamma$ denotes that the fluctuations are weighted by a vertex function) is related to the generalized structure factor $S_{\gamma \gamma}$ through the fluctuation-dissipation theorem. The inelastic response, in the long wavelength limit, is sensitive to effective density fluctuations $\left(\chi_{\gamma\gamma}(\omega)  \equiv \chi_{\gamma \gamma}(0,\omega)\right)$ given by
\begin{equation}
\chi_{\gamma \gamma}(\omega) =  \int_{0}^{\beta} d\tau e^{- i \omega_m \tau }\langle T_{\tau} \tilde{\rho}_{\gamma}(\tau),\tilde{\rho}_{\gamma}(0) \rangle |_{i \omega_m \rightarrow \omega+ i \delta}
\end{equation}
where the effective (weighted) density is written as
\begin{equation}
\tilde{\rho_{\gamma}} = \sum_{\vec{k},\sigma} \sum_{n,m} \gamma_{n,m}(\vec{k}) c_{n,\sigma}^{\dagger}(\vec{k}) c_{m,\sigma}(\vec{k}), 
\end{equation}
 $n,m$ denote band indices, $\gamma_{n,m}$ is the vertex, and $c_{n,\sigma}^{\dagger}(\vec{k})$ is the electron creation operator for band $n$, momentum $\vec k$ and spin $\sigma$. If we are interested only in the low energy response, we can neglect the off-diagonal vertices by  substituting $\gamma_{n,m} \rightarrow\delta_{mn} \gamma_n$, and thereby ignore interband transitions to write the vertex in the form
\begin{eqnarray*}
\gamma_{n}(\vec{k}) &=& \hat e^i \hat e^s  + \frac{1}{m}\sum_{j \neq n}\frac{ \langle n, \vec{k} | \hat e ^s p | j, \vec{k} \rangle \langle j, \vec{k} | \hat e^i p | n, \vec{k} \rangle}{\epsilon_{n}(\vec{k}) - \epsilon_j(\vec{k}) + \omega_I}\\
&&+\frac{ \langle n, \vec{k} | \hat e ^i p | j, \vec{k} \rangle \langle j, \vec{k} | \hat e^s p | n, \vec{k} \rangle}{\epsilon_{n}(\vec{k}) - \epsilon_j(\vec{k}) - \omega_S}.
\label{eq:Vertex}
\end{eqnarray*}
\begin{figure}[h!]
\includegraphics[width=2in,height=1.55in]{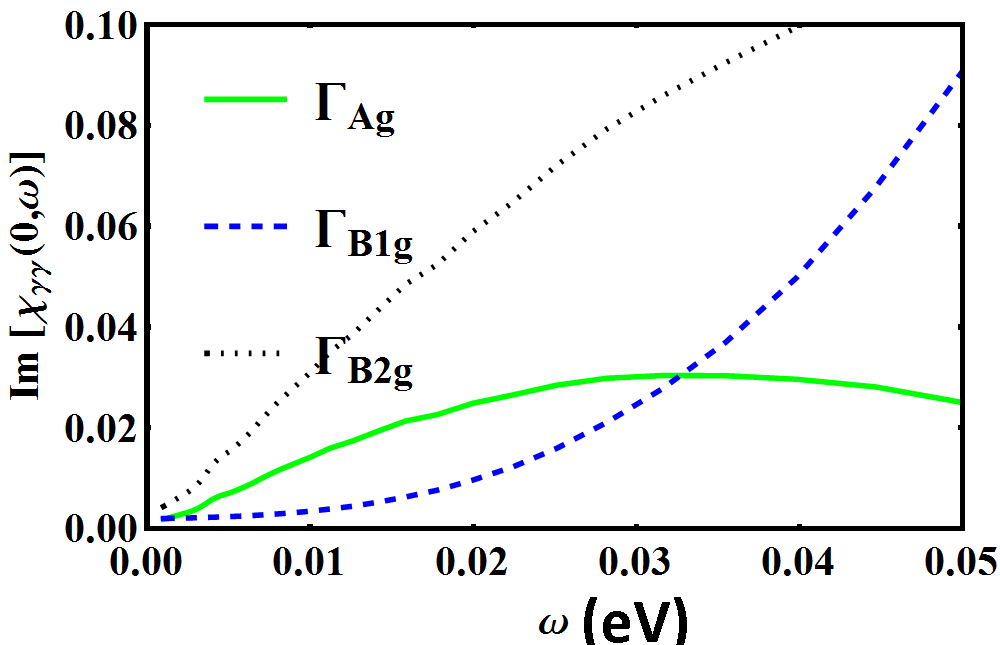}\hfill%
\includegraphics[width=1.3in,height=1.5in]{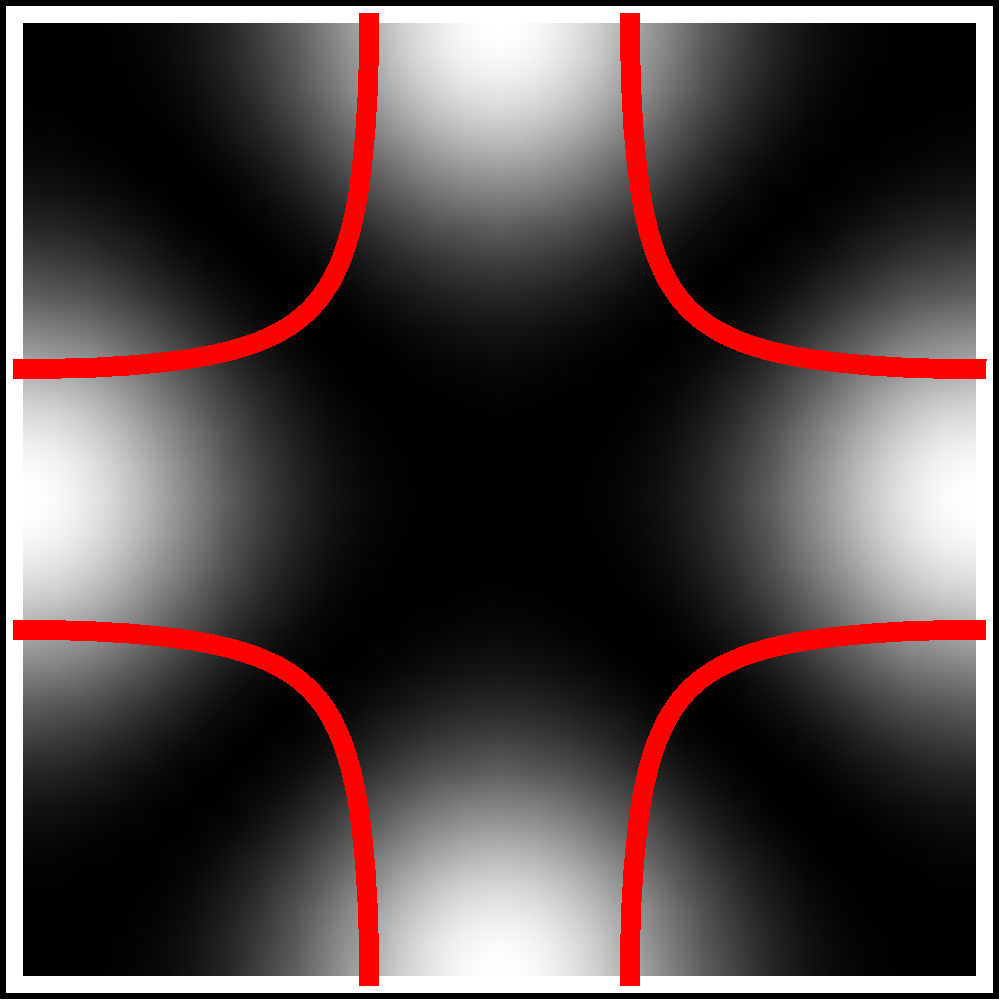}%
\caption{(Left)  Imaginary part of the low energy effective Raman response,  $\left(\chi_{\gamma\gamma}(\omega)  \equiv \chi_{\gamma \gamma}(0,\omega)\right)$, for a square lattice for different polarization geometries for the gap function $\Delta_0 (cos k_x - cos k_y)$ for $\Delta_0 = 30 meV$. (Right) Fermi surface (red contour) of a nearest and next-nearest neighbor dispersion as well as the square of the d-wave gap function plotted in the Brillouin zone. The bright (dark) regions indicate large (small) squared gap values. The function is exactly zero along the zone diagonals.  $\Gamma_i$ denote the IRRPs.}\label{d-wave}
\end{figure}
 Here we have defined the electron mass $m$, $| n, \vec k\rangle$ as the Bloch state with band index $n$ and crystal momentum $\vec k$, $\hat{e}^{i,s}$ denotes the polarization directions of the incident and scattered light respectively, $p$ is the momentum operator and $\epsilon_n(\vec k)$ are the Bloch energies.  
The vertex function on the $n$th band can be broken down into various contributions from the IRRPs of the point group as
\begin{equation}
\gamma_n(\vec{k}) = \sum_{\mu} \gamma_n^{\mu} \mathscr{I}_n^{\mu}(\vec{k})
\end{equation}  
where the index $\mu$ denotes the contributions from the different point group irreducible representations and the functions $\mathscr{I}^{\mu}_n(\vec{k})$ are the corresponding basis functions of the $\mu$'th IRRP in the $n^{\rm th}$ band. The Raman response from the $n^{\rm th}$ band can then be written as (in the independent band approximation)
\begin{eqnarray}
\chi_{\gamma\gamma}^{(n)}(i\omega) &=& \sum_{\vec{k}} |\gamma_{n}(\vec{k})|^2 \lambda_{n}(\vec{k},i\omega) 
\label{eq:ImResponse}
\end{eqnarray}
with $\lambda_n$ the Tsuneto function given by\cite{Hackl2007,Einzel1995,Neumeier1994,Zawadowski1996}
\begin{equation}
\lambda_n(\vec{k},i \omega) = \frac{\Delta_n(\vec{k})^2}{E_n(\vec{k})^2} \left( \frac{1}{2 E_n(\vec{k}) + i \omega}+\frac{1}{2 E_n(\vec{k}) - i \omega}\right),
\end{equation}
where $\Delta_n(\vec k)$ is the gap function on the $n^{\rm th}$ band and $E_n(\vec k)^2$ are the quasiparticle energies and equals $\epsilon_n(\vec k)^2+ \Delta_{n}(\vec k)^2$. As we assumed earlier, for simplicity, we will take the gap function to be independent of the orbital or band indices and denote it by $\Delta(\vec k)$. We will use the expression in Eq. (\ref{eq:ImResponse}) to evaluate the response functions in the following sections. We would like to point out that in deriving the expressions in  Eq. (\ref{eq:ImResponse}), we have ignored the role of vertex corrections through final state interactions, and long range Coulomb interactions. A recent work~\cite{maiti2017conservation} studied these effects in a generic multi-band system and concluded that vertex corrections remove the $2\Delta_0$ singularity and create a broad peak at higher energies. These conclusions should not invalidate our low energy calculations, but must only modify the exact position of the peaks.  The authors also find that in the $q\rightarrow0$ limit, long range Coulomb interactions are irrelavant to the Raman response for all polarization geometries. Additional effects arising from subleading interaction channels due to vertex corrections leading to collective in-gap ($\omega<2 \Delta_0$) Leggett and Bardasis-Schrieffer modes will not be considered in this study. We have also assumed that the light scattering occurs within a region where superconductivity is uniform. In the presence of domain walls separating different superconducting states, the responses must be averaged over the domains, and isolating the pairing symmetry is not straightforward. \\ \newline
\begin{figure}[h!]
\includegraphics[width=1.5in,height=1.35in]{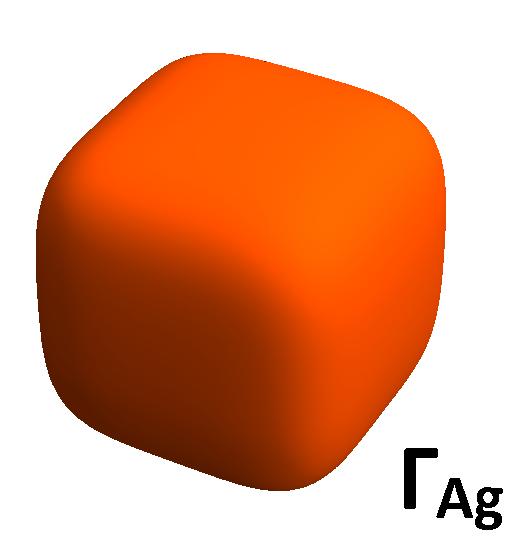}\hfill%
\includegraphics[width=1.5in,height=1.45in]{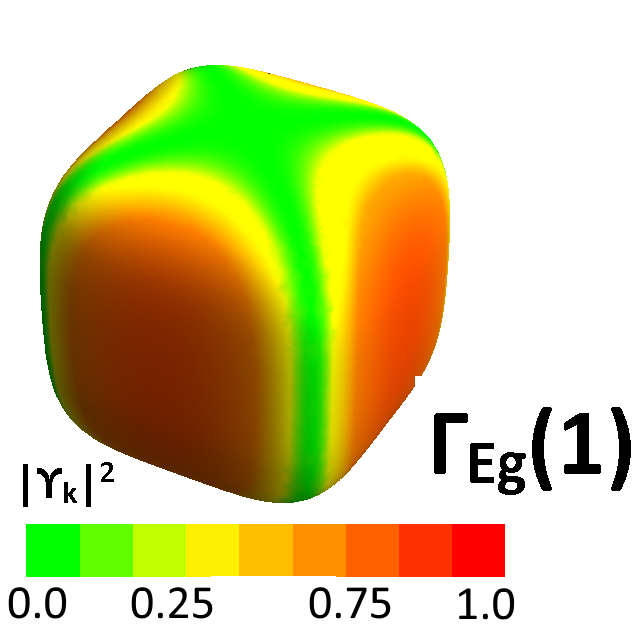}\hfill%
\includegraphics[width=1.5in,height=1.35in]{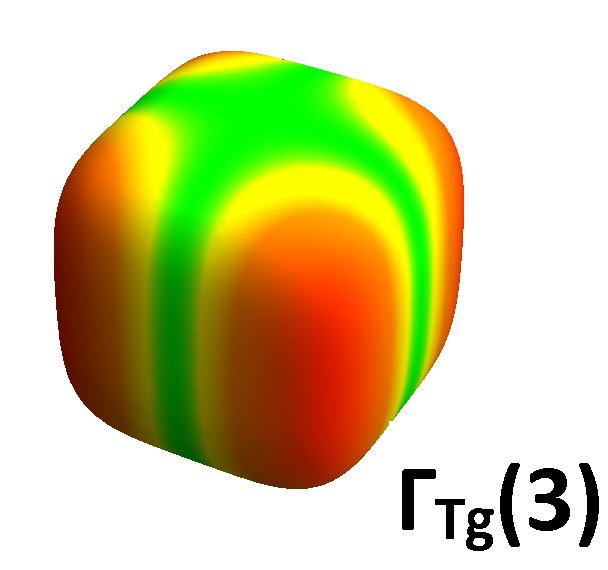}\hfill%
\includegraphics[width=1.5in,height=1.35in]{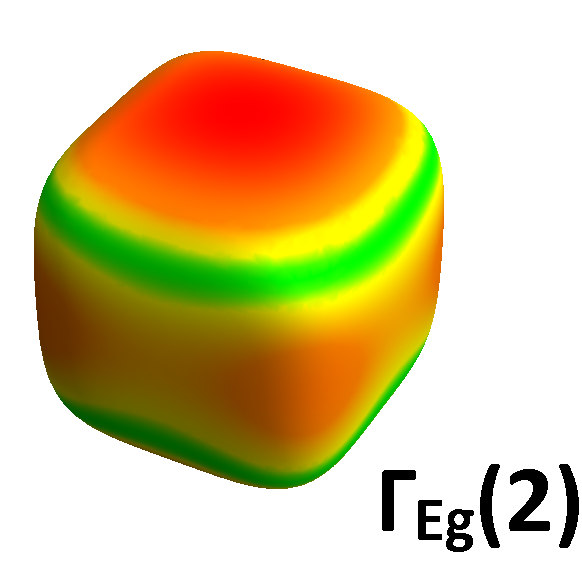}%
\caption{Square of the vertex ``weighting factor", $\gamma_{\vec{k}}^2$, plotted on the outer cubic fermi surface. Clockwise from top left: constant $\Gamma_{A_g}$ (constant), $\Gamma_{E_g}(1)$ ($x^2 - y^2$), $\Gamma_{E_g}(2) (z^2 - \frac{1}{2}(x^2+y^2))$ and $\Gamma_{T_g}(3)(xy)$ }\label{Vertex}
\end{figure}
\subsection{Light Polarization and Symmetry:\\ Square Lattice} As a warm up, we recall the arguments for the case of probing a $d$-wave gap for a square lattice (e.g. the Cuprates)\cite{Einzel1995}. We start with a dispersion of the form $\epsilon(\vec k) = - 2t( \cos(k_x) + \cos(k_y) ) + 4 t' \cos(k_x)\cos(k_y)- \mu$ where we choose $(t, t', \mu) = (1, 0.65,-1) eV$. This gives us a Fermi surface contour shown in the right panel of fig \ref{d-wave}. The crucial feature that makes inelastic light scattering of Bloch electrons powerful, is the ability of the experimentalist to control the vertex function $\gamma_n(\vec k)$ by manipulating the incoming and scattered light polarization geometries $\hat{e}^i$ and $\hat{e}^s$. For a generic multiband system (one can think of a single band system as a multiband system with infinite energy separation between bands), as long as the bands are substantially (compared to the frequency of incoming/scattered light) far apart near the Fermi energy, one can approximate the equation for the vertex function in Eq~(\ref{eq:Vertex}) as a band curvature \cite{Einzel1995,Zawadowski1996}. Note that this approximation breaks down when the bands touch each other at or near the fermi level and could yield qualitatively different results \cite{Setty2014}. Keeping in mind that there are three distinct irreducible representations, $-A_{1g}, B_{1g}, B_{2g} -$ that have been accessed in the square lattice point group, each IRRP (selected through its appropriate polarization geometry) can selectively average different regions of the Fermi surface, yielding distinct responses in the superconducting state. The basis functions for the square lattice IRRPs are given by $B_{1g}\sim \cos(k_x)-\cos(k_y)$, $B_{2g}\sim \sin(k_x)\sin(k_y)$ and $A_{1g}\sim \cos(k_x) + \cos(k_y)$. The $B_{2g}$ geometry can be accessed, for example, by setting $\hat{e}^i = (1,0), \hat{e}^s = (0,1)^T$ or vice-versa, while the $B_{1g}$ representation can be obtained by rotating the $B_{2g}$ polarization vectors by $\pi/4$. To access the $A_{1g}$ representation, one needs to use the left and right circularly polarized light for the incoming/outgoing polarization vectors. The low energy inelastic response in these geometries for the square lattice is plotted in Fig \ref{d-wave}(left) for a $d-$wave gap function of the form $\cos(k_x) - \cos(k_y)$. The low energy curves behave very anisotropically for different geometries.  For example, in the $A_{1g}$ and $B_{2g}$ cases, they behave linearly for small $\omega$ while for the $B_{1g}$ case, the response is super-linear ($\sim \omega^3$) \cite{Einzel1995,Hackl2007}. To understand this behavior one has to look at the regions of the Fermi surface that are being sampled$-$ in the $B_{2g}$ geometry, the momentum average is focussed along the $k_x = \pm k_y$ directions where the $d-$wave nodes are located, while the $B_{1g}$ geometry samples the anti-nodal regions (along the $k_x$ or $k_y$ axes) of the Fermi surface. Thus the low energy response in the $B_{1g}$ grows slower ($\sim \omega^3$) relative to the $B_{2g}$ case ($\sim \omega$). In the $A_{1g}$ scenario, there is sampling of both the nodal and anti-nodal regions and, thus, the low energy response is dominated by the nodes ($\sim \omega$). In the following paragraphs, we will follow a similar line of argument for the $T_h$ point group.\\ 
 \newline
\subsection{Light polarization and symmetry$-T_h$ point group} In this section, we will highlight the connections between light polarizations that can be experimentally manipulated, and the symmetry of the Raman vertex in the Brillouin zone for the $T_h$ point group. For simplicity, we will illustrate this connection using dispersions obtained by individual IRRPs for the calculation of the Raman vertex. The total vertex for the band structures (used above) can then be obtained by a simple linear combination of the vertices of these representations. In order to keep our equations tractable in this example, we will work in the continuum limit; this can be easily generalized to the lattice case with the inclusion of trigonometric functions in momenta instead of polynomials. Let us begin with a dispersion obtained from the $\vec{E}_g$ representation appearing as a term in Eqs. \ref{Dispersions}. In the continuum limit, this is given by
\begin{eqnarray}
\epsilon_{E_g}(\vec k) &=& \vec{E}_{g}(\vec k)\cdot\vec{E}_g(\vec k) - \mu \\
\vec{E}_{g}(\vec k) &=& \left( (k_x^2- k_y^2) , 2 k_z^2 - (k_x^2 + k_y^2) \right),
\end{eqnarray}
with $\mu$ the chemical potential. We can evaluate the Raman vertex for this dispersion using the band effective mass approximation for the $n^{\rm t}$ band as\cite{Zawadowski1996}
\begin{eqnarray}
\gamma_n(\vec k) &=& \sum_{\mu\nu} \hat{e}_{\mu}^i R^{\mu\nu}_{\Gamma} \hat{e}^s_{\nu},\\
R^{\mu\nu}_{\Gamma} &=&m \frac{\partial^2 \epsilon_{\Gamma}(\vec k)}{\partial k_{\mu} \partial k_{\nu}}
\end{eqnarray}
where,  $\epsilon_{\Gamma}(\vec k)$ is the band dispersion formed from the irreducible representation $\Gamma$ and $\mu, \nu$ are the coordinate indices. For these dispersions, the  Raman tensor $R^{\mu \nu}$  is 
\begin{equation}
R^{\mu \nu}_{E_g} = 
 \begin{pmatrix}
 3 k_x^2 - k_z^2& 0  &-2 k_x k_z \\
0 & 3 k_y^2 - k_z^2  & - 2 k_y k_z \\
 - 2 k_x k_z& -2 k_y k_z &  - k_x^2 - k_y^2 + 6 k_z^2
 \end{pmatrix}.
\end{equation} 
An arbitrary choice of the incoming and scattered polarization vectors yields a combination of $A_g$, $\vec{E}_g$ and $\vec{T}_g$ representations. We can obtain a pure representation by choosing $\hat{e}^i$ and $\hat{e}^s$ such that only the representation of interest contributes to the scattering cross section. This is not always straightforward, as we will see, and one will need to resort to subtraction procedures. The simplest cases to access are the $E_g(1)$ and $\vec{T}_{g}$ representations. Choosing $\hat{e}^i = (1,-1,0)^T$ and $\hat{e}^s = (1,1,0)$, it is easy to verify that $ \sum_{\mu\nu} \hat{e}_{\mu}^i R^{\mu\nu}_{\Gamma} \hat{e}^s_{\nu} = 3 (k_x^2 - k_y^2)$, which transforms as the first component of the $\vec{E}_g$ representation. Similarly, choosing $\hat{e}^i = (0,0,1)^T$ and $\hat{e}^s = (1,0,0)$, or, $\hat{e}^i = (0,0,1)^T$ and $\hat{e}^s = (0,1,0)$, picks up the $T_g(1)$ or $T_g(2)$ representations respectively (the $T_g(3)$ component is zero here). To obtain the $E_g(2)$ and $A_g$ representations, we need to choose two different sets of incoming and scattered polarizations.  As our first choice, ($c_+$), we select $\hat{e}^i = (\sqrt{\alpha},i \sqrt{\alpha},\sqrt{\beta})^T$ and $\hat{e}^s = (\sqrt{\alpha},-i \sqrt{\alpha},\sqrt{\beta})$, and for our second choice ($c_-$), we select $\hat{e}^i = (\sqrt{\alpha},i \sqrt{\alpha},-\sqrt{\beta})^T$ and $\hat{e}^s = (\sqrt{\alpha},-i \sqrt{\alpha},-\sqrt{\beta})$. Here, $\alpha$ and $\beta$ are constants that must be determined. These two choices $c_{\pm}$ for the polarization give us vertices
\begin{eqnarray}
\label{gamma-pm}
\gamma_n(\vec k)_{\pm} &=& (3\alpha - \beta)\left( k_x^2 + k_y^2\right) + (6\beta - 2 \alpha) k_z^2 \\ \nonumber
&&  \pm \sqrt{\alpha\beta} (-2 k_x k_z).
\end{eqnarray}  
At this point, the constants $\alpha$ and $\beta$ can be related to each other in two useful ways. We consider first the case where the coefficients of $(k_x^2+ k_y^2)$ and $k_z^2$ are equal, i.e, when $(3\alpha - \beta) = (6\beta - 2\alpha)$ or $5\alpha = 7\beta$. With this condition, the vertices $\gamma_n(\vec k)_{\pm}$ in Eq~\eqref{gamma-pm} are the sum and difference of the $A_g$ ($k_x^2 + k_y^2 + k_z^2$) and $T_g(1)$ ($k_x k_z$) representations alone. Thus, we have
\begin{equation}
\gamma_n(\vec k)_{\pm} = \gamma(\vec k)_{A_g }\pm \gamma(\vec k)_{T_g(1)},
\end{equation}
where  $\gamma(\vec k)_{A_g }$ and $\gamma(\vec k)_{T_g(1)}$ are vertices that individually transform as $A_g$ and $T_g(1)$ representations respectively. Summing the responses from $\gamma_n(\vec k)_+$ and $\gamma_n(\vec k)_-$ using Eq. ( \ref{eq:ImResponse}), and noting that we know how to obtain a pure $T_g(1)$ contribution, we can calculate the response from a pure $A_g$ representation by a mere subtraction of the two resulting responses. For the pure $E_g(2)$ response, we equate the coefficients of the $(k_x^2+ k_y^2)$ and $k_z^2$: $6\beta - 2 \alpha = -2(3\alpha - \beta)$ or $\alpha = - \beta$. Substituting this condition into Eq. (\ref{gamma-pm}), we obtain the vertices as 
\begin{equation}
\gamma_n(\vec k)_{\pm} = \gamma(\vec k)_{E_g (2)}\pm i \gamma(\vec k)_{T_g(1)},
\end{equation}
where $\gamma(\vec k)_{E_g (2)}$ is the vertex that transforms as the second component of the $E_g$ representation. Again using Eq. (\ref{eq:ImResponse}), for either $\gamma_n(\vec k)_{\pm}$, and equipped with knowledge of the $T_{g}(1)$ response, we can isolate the response in the $E_g(2)$ channel by subtracting the two resulting responses.\\ \newline
As an example of a dispersion that can be obtained through an $A_g$ representation in Table \ref{Character}, we choose a dispersion of the form
\begin{equation}
\epsilon_{A_g}(\vec k) = c_x + c_y + c_z + c_x c_y + c_y c_z + c_x c_z- \mu, \\
\end{equation}  \\ \newline
which is a generalization of the function $A_g(\vec k)$ used in eq \ref{Irreps}. The Raman tensor $R^{\mu \nu}$ for this dispersion is given as
\begin{equation}
R^{\mu \nu}_{A_g} = 
 \begin{pmatrix}
 -C_{x; yz}& s_x s_y  & s_x s_z \\
s_x s_y & -C_{y; zx} & s_y s_z \\
 s_x s_z& s_y s_z &  - C_{z;xy},
 \end{pmatrix}.
\end{equation} 
where we have defined $C_{i;jk} = - c_i (1+ c_j + c_k)$. To obtain the components of a pure $\vec T_{g}$ representation, we can follow arguments analogous to what we did in the previous paragraph. For the third component of the $\vec T_{g}$ representation, we need to choose $\hat{e}^i = (0,1,0)^T$ and $\hat{e}^s = (1,0,0)$, and similarly for the first and second components. To obtain the response in the $A_g$ channel, we choose $c_+$ to be $\hat{e}^i = (1,i,1)^T$ and $\hat{e}^s = (1,-i,1)$ and select $c_-$ to be $\hat{e}^i = (1,i,-1)^T$ and $\hat{e}^s = (1,-i,-1)$. The resulting vertices corresponding to these two choices are the sum and difference of the $A_g$ and $T_g(1)$ symmetries; the pure $A_g$ response can then be extracted by using eq \ref{eq:ImResponse} for these vertices, along with our knowledge of the pure $T_g(1)$ response. One can repeat this analysis for a dispersion $\epsilon_{T_u}(\vec k)$ (of the form appearing in eq \ref{Dispersions}) obtained from the basis functions of the $\vec T_{u}$ representation in Table \ref{Character}. Fig \ref{Vertex} shows regions of the outer hole pocket that are sampled by the various representations studied in this section. We will use this figure to understand the frequency dependence of the low energy inelastic response in the following section. \\ \newline
At this point, it is important to note certain circumstances where this procedure of extracting pure irreducible representations breaks down.  First, in order to isolate the cross section from a pure irreducible representation, we note that the subtraction process requires some basic assumptions about the band structure and superconducting gaps. Such assumptions have been made in the context of the cuprates as well \cite{Hackl2007,Einzel1995,Neumeier1994,Zawadowski1996}.  In our analysis, we used simple band structures created from invariants using basis functions in Table \ref{Character}; but once a more accurate tight-binding band structure calculation emerges, our subtraction procedure can be used to analyse inelastic Raman data with greater confidence.  More complications could arise when there are multiple bands crossing the Fermi level (as the case is here) due to the fact that a pure IRRP in one band need not correspond to the same IRRP in the other. In such cases, one could still think of indirect ways to avoid mixing of different representations. For example, on could rely on the fact that, in general, the gaps on different bands have different magnitudes. With an approximate knowledge of the values of the gaps on each band (from experiments like ARPES), it is possible to isolate the response from an individual band by keeping only the response in the relevant energy window and subtracting the contributions not in that window; this could be done for each polarization geometry separately. Although this procedure is less straighforward, one could avoid mixing of different IRRPs to a certain extent. However, this procedure becomes ambiguous when the gaps are similar in magnitude and/or have different form factors on different bands, and could have limited applicability. In such cases, one has to be satisfied with studying the responses with multiple contributing IRRPs (all be it with knowledge of what IRRPs form the combination). One could still make reasonably accurate comparisons with the corresponding theoretical predictions of the responses in mixed geometries Second, in the presence of vertex corrections, there are additional contributions that need to be incorporated into the inelastic response \cite{Zawadowski1996, maiti2017conservation} which might, in general, make it difficult to isolate pure representations. Third, the effective mass approximation used in our analysis is valid only when the frequency of the incoming and scattered light is either small or  large in comparison to the inter-band spacings \cite{Zawadowski1996,Setty2014}.
\begin{figure}[h!]
\includegraphics[width=2.1in,height=1.5in]{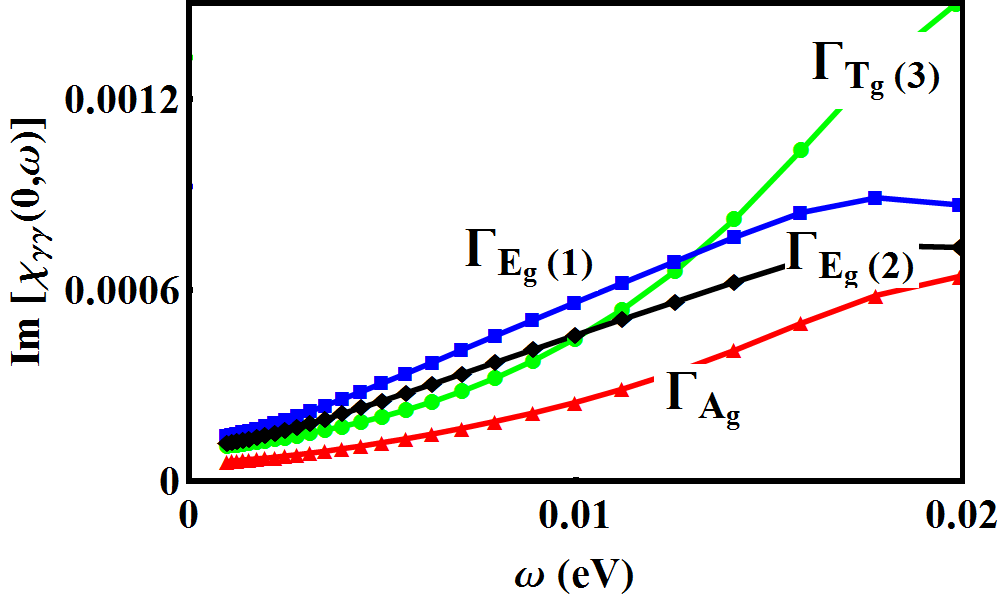}\hfill%
\includegraphics[width=1.3in,height=1.55in]{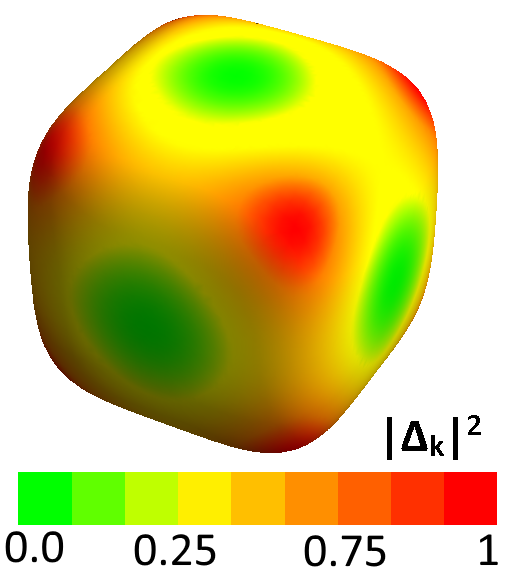}%
\caption{(Left)Imaginary part of the low energy effective Raman response,  $\left(\chi_{\gamma\gamma}(\omega)  \equiv \chi_{\gamma \gamma}(0,\omega)\right)$, in POS for different polarization geometries for the gap function $ \Delta(\vec k) = \Delta_0 (k_x^2 k_y^2 + k_y^2 k_z^2 +  k_z^2 k_x^2)$ with $\Delta_0 = 0.5 eV$.  Note that $\Delta_0$ has to be large to get an experimentally reasonable value of the gap due to the fourth powers in momentum. (Right) the outer (cube-like) fermi hole pocket centered around the $\Gamma$ point. The color scale on the Fermi surface denotes the value of the gap function $\Delta(\vec k) = \Delta_0 \Phi_s(\vec k)$ that has six nodes, two along each of the coordinate axes. The gap has been normalized such that its maximum value on the Fermi surface is 1. }\label{Delta6Dip}
\end{figure}
 This may not always be accurate \cite{Setty2014} and one must be careful in not overestimating the vertex in certain regions of the Fermi surface. Due to the above outlined reasons, our calculations should not be considered any more than qualitative.\\ \newline
\subsection{Low energy Raman response} Having outlined the procedure to isolate the response from select irreducible representations above, we now demonstrate its use in determining the exact form of the pairing symmetry in POS. We confine ourselves to the spin singlet, even parity forms of the gap functions proposed in Goryo's work \cite{Goryo2003}. Similar arguments can be carried out in a straightforward manner for other pairing forms. We choose the temperature and broadening parameter in the response function to be 2 meV and 3 meV respectively. Fig \ref{Delta6Dip} (right) shows a color plot of the $A_g$ gap function, $\Delta(\vec k) = \Delta_0 \Phi_s(\vec k)$, proposed in the A phase of POS. The brighter (green) regions show the location of the six nodes on each face of the cube. Fig \ref{Delta6Dip} (left) shows the low energy response for such a gap function in channels belonging to different irreducible representations. A suppression of spectral weight for small $\omega$ is observed in the $T_g(3)$ ($xy$) geometry, with the $E_g(1)$ and $E_g(2)$ geometries showing enhancement for small $\omega$. To understand this result, we see from Fig \ref{Vertex} that the Fermi surface averaging in the $T_g(3)$ channel occurs mainly along the $x=\pm y$ directions on the $x-y$ plane. It is exactly along these directions where the chosen $A_g$ gap function makes the Fermi surface fully gapped, thus, suppressing low energy quasiparticle response. The $E_g(1)$ and $E_g(2)$ geometries, on the other hand, sample at least two of the gap nodes (see Fig \ref{Vertex}) and excite a low energy quasi-particle response. To confirm this picture, we repeated the calculation for a four-dip state given by $\Delta(\vec k) = \Delta_0( k_x^2 k_y^2 + k_y^2 k_z^2 + a k_z^2 )$, which has nodes only along the $k_x$ and $k_y$ axes. In this case, in addition to the $T_g(3)$ geometry, a small suppression of quasiparticle spectral weight occurs in the $\Gamma_{E_g}(2)$ geometry as well. The suppression is only small due to the fact that the $E_g(2)$ geometry $-$inspite of its extended lobe-like feature along the $k_z$ axis$-$ has a small vertex contribution coming from the equatorial plane as well (see fig~\ref{Vertex} bottom right). \\ \newline
\begin{figure}[h!]
\includegraphics[width=2.2in,height=1.6in]{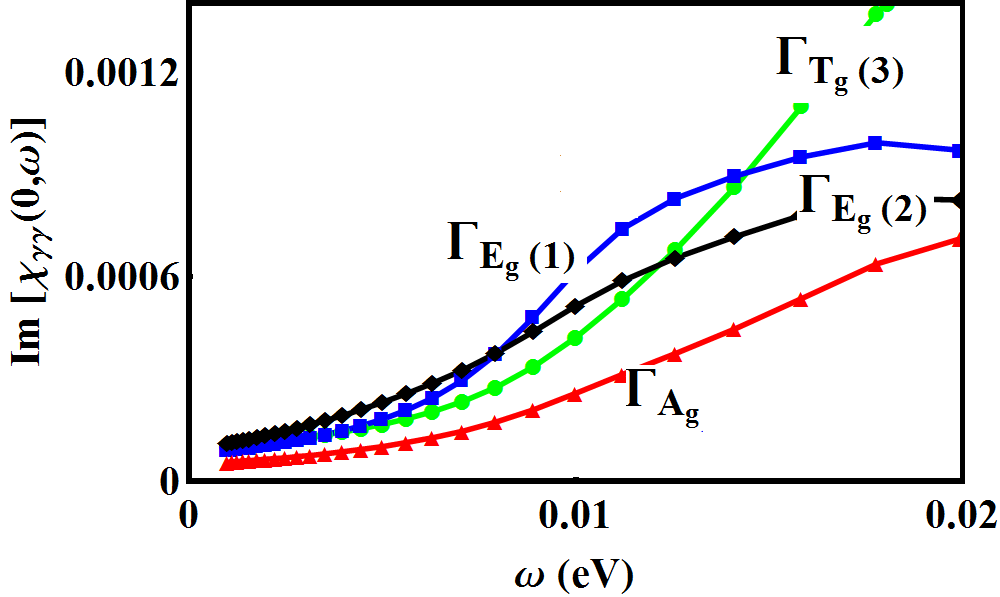}\hfill%
\includegraphics[width=1.2in,height=1.5in]{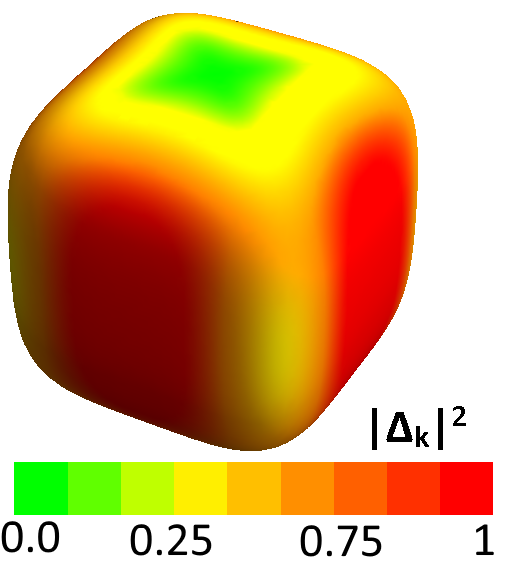}%
\caption{Same as Fig \ref{Delta6Dip} but with a gap function that has only two nodes instead as observed in \cite{Maki2003}. Because the point group of POS has a $T_h$ symmetry, we can arbitrarily choose the two nodes to lie along the $k_z$ axis with no nodes along $k_x$ and $k_y$. (Left) Imaginary part of the low energy effective Raman response in POS for different polarization geometries for the gap function $ \Delta(\vec k) = \Delta_{s0}(k_x^2 k_y^2 + k_y^2 k_z^2 +  k_z^2 k_x^2) + i \Delta_{d0} (k_x^2 -  k_y^2)$ $ \left(\equiv \left( \Phi_s(\vec k)  +  i \Phi_d(\vec k) \right)\right)$ with $\Delta_{s0} = 0.5eV$ and $\Delta_{d0} = 50 meV$.  Note that $\Delta_{s0}$ has to be large to get an experimentally reasonable value of the gap due to the fourth powers in momentum. (Right) The color function in this panel corresponds to the absolute value of the gap, as measured by Raman scattering. The gap has been normalized such that its maximum value on the Fermi surface is 1.}\label{TRSB}
\end{figure}
We can follow a similar analysis in the B phase. For the sake of convenience, we choose a gap function, $\Delta(\vec k) =  \left( \Phi_s(\vec k)  +  i \Phi_d(\vec k) \right) \equiv  \Delta_{s0}(k_x^2 k_y^2 + k_y^2 k_z^2 +  k_z^2 k_x^2) + i \Delta_{d0} ( k_x^2 - k_y^2)$, such that there are two nodes along the $k_z$ axis (such a choice is possible provided the basis functions are also transformed appropriately). Given that the intensity of the Raman cross section depends only on the absolute value of the gap function, it is insensitive to the total phase of the order parameter but is sensitive to relative phases of the individual components. Fig (\ref{TRSB}) (right) shows the gap nodal structure in the B phase $-$ the Fermi surface now has nodes along the $k_z$ direction and is  gapped everywhere else. Following arguments presented in the previous paragraph, we see that in the $T_{g}(3)$ and $E_{g}(1)$ geometries, which represents sample regions primarily along the $k_x = \pm k_y$  and $k_x=0, k_y=0$ directions respectively, the low energy response is reduced (see Fig (\ref{TRSB}) left panel, and Fig~\ref{Comparison} which compares the $E_g(1)$ response for different pairing forms). However, in the $E_g(2)$ geometry, the gap function along the $k_z$ axis is projected; as a result, one should expect an increased low energy response. In the $A_g$ geometry, all the high symmetry regions are averaged over and results in a response intermediate in curvature to the other geometries. The four-dip state in the TRSB has been argued \cite{Goryo2003} to be highly accidental and hence we will not discuss this state. However, if indeed, this state were to be realized, one would obtain a low energy increase of the intensity in the $\Gamma_{E_g}(1)$ geometry as well. \\ \newline
\begin{figure}[h!]
\includegraphics[width=2.4in,height=1.7in]{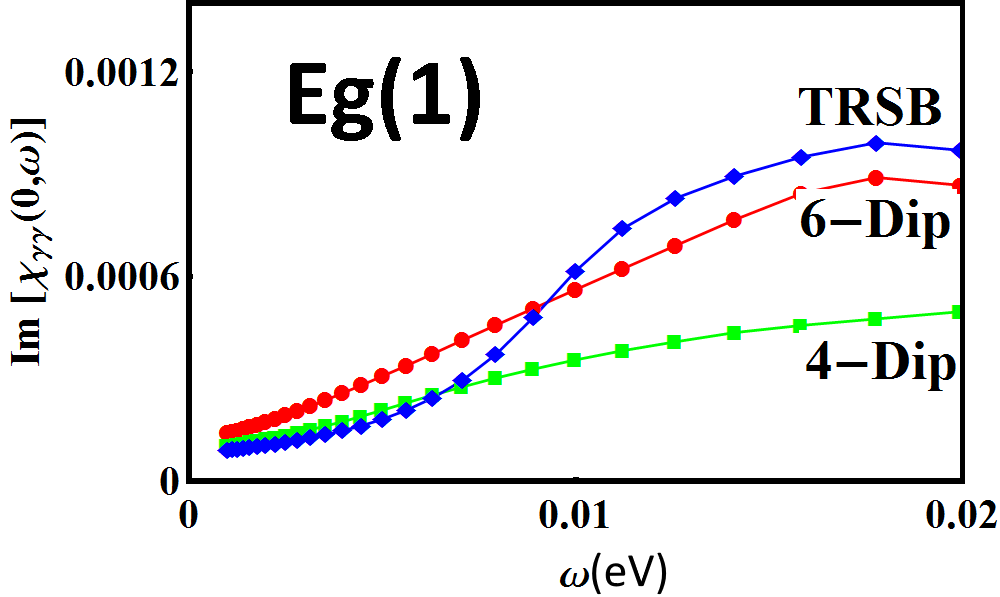}\hfill%
\caption{Comparison of the low energy responses,  $\left(\chi_{\gamma\gamma}(\omega)  \equiv \chi_{\gamma \gamma}(0,\omega)\right)$, in the $E_g(1)$ for different pairing form factors as marked. There is a clear suppression (slower than linear behavior) of the
spectral weight for the TRSB case in the energy scale of the order of the gap. The 4-Dip state is described by $\Delta(\vec k) = \Delta_0( k_x^2 k_y^2 + k_y^2 k_z^2 + a k_z^2 )$ where $a=0.5 $ and $\Delta_0 = 0.4 eV$ and has four nodes all lying on the equatorial plane. The 6-Dip state is described by $\Delta(\vec k) = \Delta_0( k_x^2 k_y^2 + k_y^2 k_z^2 +  k_z^2 k_x^2 )$ ($\Delta_0 = 0.5$ eV) with four nodes along the equator and two along the axial direction. TRSB corresponds to the time reversal symmetry breaking state with only two nodes along the axial direction. }\label{Comparison}
\end{figure}
At this juncture, we would like to emphasize what we briefly pointed out earlier regarding the response in the $E_{g}(2)$ geometry. For the gap with four nodes, this geometry exhibits an intermediate behavior between enhancement and suppression. This is because there is also an equatorial contribution, although small, that cannot be entirely neglected compared to the axial contribution (see Fig \ref{Vertex}). One should, therefore, in the process of trying to extract nodal behavior in gap functions, interpret the $E_g (2)$ response with care.\\ \newline
\section{Final Remarks} We derived a Ginzburg-Landau theory starting from a microscopic model that describes the entry of the skutterudite superconductor  $\mathrm{PrOs_{4}Sb_{12}}$ (POS) into a time reversal symmetry broken (TRSB) phase as a function of temperature, in accordance with recent AC susceptibility and polar Kerr effect measurements \cite{Kapitulnik2016}. Using the expansion, we calculated the GL coefficients using their relationships to the various microscopic parameters and determined the shape of the contour bounding the TRSB phase. As an experimental means to probe the proposed pairing symmetries, we examined the low energy inelastic (Raman) response in both the A and B phases of POS. We provided a  qualitative understanding of the low energy and zero momentum transfer response for various light polarization geometries. By appropriate manipulation of the incoming and scattered light geometries, as well as other subtraction procedures, we demonstrated that one can access the various irreducible representations contained in the $T_h$ point group describing POS. Depending on the existence of nodes along the $c-$ axis, we found an enhancement (nodal) or suppression (gapped) of the low energy spectral weight, based on what regions of the Fermi surface were being sampled. Inelastic light scattering could, thus, help pin-point the exact location of nodes on the fermi surface and, thereby, narrow down the candidate pairing symmetries proposed in POS.\\ \newline
\textit{Acknowledgments:} CS and PWP acknowledge support from Center for Emergent
Superconductivity, a DOE Energy Frontier Research Center, Grant No. DE-AC0298CH1088. YW is supported by the Gordon and Betty Moore Foundations EPiQS Initiative through Grant No. GBMF4305.
\bibliographystyle{apsrev4-1}
\bibliography{PrOsSb}
\onecolumngrid 
\newpage
\section{Appendix:  Landau-Ginzburg Analysis}
Here we outline some details about calculations involved in obtaining the phase diagrams appearing in Fig 3 of the main text. We will start by repeating the following form of the action written in momentum space:
\begin{eqnarray}\nonumber
S(\bar c, c) &=& \beta \sum_{ \substack{\alpha\beta \\ \textbf k \sigma }} \bar c_{\alpha \sigma}(\textbf k)\left(- ik_n\delta_{\alpha\beta} - \epsilon_{\alpha \beta}(\vec k)\right) c_{\beta \sigma}(\textbf k) \\ 
&& - \beta g \sum_{\substack{\textbf k \textbf k'\\ \textbf q \alpha}} \bar c_{\alpha \uparrow}(\textbf k + \textbf q)  \bar c_{\alpha \downarrow}(\textbf k' - \textbf q) \tilde{V}(\vec k, \vec q)  c_{\alpha \downarrow}(\textbf k')  c_{\alpha \uparrow}(\textbf k).  
\end{eqnarray}
Here  $\bar c_{\alpha \sigma}(\textbf k)$ and  $c_{\beta \sigma}(\textbf k)$ are Grassmann numbers that follow Grassmannian algebra for electrons with Bloch momentum $\vec k$ and Matsubara frequency $k_n$ (collectively denoted by $\textbf{k}$), spin $\sigma$ and orbital indices $\alpha, \beta$, $\epsilon_{\alpha \beta}(\vec k )$ are the orbital matrix elements, $\beta$ is the inverse temperature (not to be confused with the index $\beta$), $g$ is the interaction strength which is taken to be a constant, and $\tilde{V}(\vec k, \vec q)$ is the four body interaction giving rise to superconductivity.  Changing to the band basis with energy eigenvalues $\epsilon_{\alpha}(\vec k)$ by using the unitary transformation, $c_{\alpha\sigma}(\textbf k) = U_{\alpha\beta}(\vec k)\psi_{\beta \sigma}(\textbf k)$, gives us the action
\begin{eqnarray}\nonumber
S(\bar{\psi},\psi) &=& \beta \sum_{ \substack{\alpha\\ \textbf k \sigma }} \bar {\psi}_{\alpha \sigma}(\textbf k)\left(- ik_n - E_{\alpha}(\textbf k)\right) \psi_{\alpha \sigma}(\textbf k) \\ 
&& - \beta g \sum_{\substack{\textbf k \textbf k'\\ \textbf q \alpha}} \bar{\psi}_{\alpha \uparrow}(\textbf k + \textbf q)  \bar{\psi}_{\alpha \downarrow}(\textbf k' - \textbf q) V(\vec k, \vec q) \psi_{\alpha \downarrow}(\textbf k')  \psi_{\alpha \uparrow}(\textbf k),
\end{eqnarray}
where $V(\vec k, \vec q)$ is the resulting interaction related to $\tilde{V}(\vec k, \vec q)$ through the band matrix elements. As we are interested in zero-momentum pairing, we will assume that $\textbf k' = -\textbf k$. After making the shift $\textbf q \rightarrow \textbf q - \textbf k$, we obtain the standard BCS action for independent bands
\begin{eqnarray}\nonumber
S(\bar{\psi},\psi) &=& \beta \sum_{ \substack{\alpha \\ \textbf k \sigma }} \bar {\psi}_{\alpha \sigma}(\textbf k)\left(- ik_n - E_{\alpha}(\textbf k)\right) \psi_{\alpha \sigma}(\textbf k) \\ 
&& - \beta g \sum_{\substack{\textbf k \textbf q \alpha}} \bar{\psi}_{\alpha \uparrow}(\textbf q)  \bar{\psi}_{\alpha \downarrow}(- \textbf q) D(\vec q)D(\vec k)  \psi_{\alpha \downarrow}(-\textbf k)  \psi_{\alpha \uparrow}(\textbf k),
\end{eqnarray}
where the interaction has been factorized in terms of $D(\vec k) \equiv \phi_s(\vec k) + \phi_d(\vec k)$, which is the total form factor and is a mixture of the $s-$ wave and $d-$wave form factors.
The total partition function can be written as
\begin{eqnarray}
\mathscr{Z} = \int \mathscr{D}(\psi,\bar{\psi}) e^{-S(\psi,\bar{\psi})}.
\end{eqnarray}
We can now decouple the quartic term appearing in the partition function using the Hubbard-Stratonovich transformation for fermions given by
\begin{equation}
\pi g e^{gab}  = \int d\phi d\bar{\phi}\hspace{3mm} e^{a \phi + b \bar{\phi}} e^{\frac{-| \phi|^2}{g}}
\end{equation}
where $a$ and $b$ are two commuting elements of a Grassmann algebra, defined by the numbers $\psi, \bar{\psi}$ which follow the anti-commutation relations $\{ \psi_i,\psi_j\} = \{ \psi_i,\bar{\psi}_j \}=\{\bar{\psi_i},\bar{\psi_j }\}=0$ with $i$ and $j$  sets of electron quantum numbers. In our case, we have defined $a\equiv\bar{\psi}_{\alpha\uparrow}(\textbf q)\bar{\psi}_{\alpha\downarrow}(-\textbf q)D(\vec q)$ and $b\equiv\psi_{\alpha\downarrow}(-\textbf k) \psi_{\alpha\uparrow}(\textbf k)D(\vec k)$ and it is easily seen that $a$ and $b$ commute. Using this, we can rewrite the partition function as
\begin{eqnarray}\nonumber
\mathscr{Z} &=& \int \mathscr{D}(\psi,\bar{\psi})\mathscr{D}(\phi,\bar{\phi}) exp\left[ -\frac{\beta| \phi|^2}{g}-\beta \sum_{\substack{\textbf{k}\alpha  \sigma}} \bar{\psi}_{\alpha\sigma}(\textbf{k})\left(-i k_n - E_{\alpha}(\vec k)\right)\psi_{\alpha\sigma}(\textbf k)\right]\\
&&\times exp\left[\beta\sum_{\substack{\textbf{k}\alpha}}\left(\phi \bar{\psi}_{\alpha\uparrow}(\textbf k) \bar{\psi}_{\alpha\downarrow}(-\textbf k) + \phi^*\psi_{\alpha\downarrow}(-\textbf{k}) \psi_{\alpha\uparrow}(\textbf k)\right)D(\vec k) \right],
\end{eqnarray}
which can be further simplified using matrix notation as
\begin{equation}
\mathscr{Z} = \int \mathscr{D}(\psi,\bar{\psi}) \mathscr{D}(\phi,\bar{\phi}) exp\left[-\beta\left(\frac{| \phi |^2}{g} + \sum_{\textbf k \alpha} \hat{\Psi}^{\dagger}_{\textbf k \alpha} \hat{\mathscr{G}}^{-1}_{\textbf k \alpha}\hat{\Psi}_{\textbf k \alpha} \right)\right].
\end{equation}
Here we have defined
\begin{equation}
\hat{\mathscr{G}}^{-1}_{\textbf k \alpha} = 
 \begin{pmatrix}
 - \left( ik_n + E_{\alpha}(\vec k)\right)&  -D(\vec k) \phi \\
-D(\vec k) \phi^* &  - \left( ik_n -  E_{\alpha}(\vec k)\right)
 \end{pmatrix}
\end{equation}
along with $\hat{\Psi}_{\textbf k \alpha} = \left(  \bar{\psi}_{\alpha \uparrow}(\textbf k),\psi_{\alpha \downarrow}(-\textbf k). \right)$. Performing the Fermionic functional integral, we obtain
\begin{equation}
\mathscr{Z} = \int \mathscr{D}(\phi,\bar{\phi}) exp\left[\frac{-\beta |\phi|^2}{g} + \sum_{\textbf k \alpha} Tr\left( Log\left[ \hat{\mathscr{G}}_{\textbf k \alpha}^{-1}\right] \right)\right].
\end{equation}
In order to perform an expansion in powers of the superconducting order parameter, we write the total Green function inverse as a sum of non-interacting and  superconducting self energy parts; that is, we write $\mathscr{G}_{\textbf k \alpha}^{-1}  = \hat{\mathscr{G}}_{0\textbf k \alpha}^{-1}  + \hat{\Sigma}$, where,
\begin{equation}
\hat{\mathscr{G}}^{-1}_{0\textbf k \alpha} = 
 \begin{pmatrix}
 - \left( ik_n + E_{\alpha}(\vec k)\right)& 0\\
0 &  - \left( ik_n -  E_{\alpha}(\vec k)\right)
 \end{pmatrix}
\end{equation}
and 
\begin{equation}
\hat{\Sigma} = 
 \begin{pmatrix}
 0&  -D(\vec k) \phi\\
 -D(\vec k) \phi^* & 0.
 \end{pmatrix}
\end{equation}
We now perform an expansion keeping in mind that $ Log\left[ \hat{\mathscr{G}}_{\textbf k \alpha}^{-1}\right] =  Log\left[ \hat{\mathscr{G}}_{0\textbf k \alpha}^{-1} + \hat{\Sigma} \right] = Log\left[ \hat{\mathscr{G}}_{0\textbf k \alpha}^{-1} \right]  + \sum_{n=1}^{\infty} \frac{(-1)^{n+1}}{n} \left( \hat{\mathscr{G}}_{0\textbf k \alpha} \hat{\Sigma} \right)^n$. All the powers odd in $n$ vanish because of the Matsubara sums and we are only left with even powers in $n$. We will only be interested in second and fourth order powers of $\Sigma$ in the Landau-Ginzburg expansion. With this, the superconducting contributions to the free energy become
\begin{equation}
\mathscr{F}_s \simeq  \frac{| \phi |^2}{g} +  \frac{1}{\beta}\sum_{\textbf k \alpha}\left(\frac{1}{2}Tr \left[\left(\hat{\mathscr{G}}_{0\textbf k\alpha} \hat{\Sigma}\right)^2\right] + \frac{1}{4}Tr \left[\left(\hat{\mathscr{G}}_{0\textbf k\alpha} \hat{\Sigma}\right)^4 \right] \right).
\end{equation}
We decompose the pairing terms into two channels: the $s-$ wave and $d-$ wave symmetries according to the point group symmetries of the lattice. For the second order contribution, we have
\begin{equation}
\mathscr{F}_s^{(2)} = T\sum_{\textbf k \alpha}\frac{|\phi_s + \phi_d|^2}{\left(E_{\alpha}(\vec k)^2 + k_n^2\right)},
\end{equation}
and the fourth order contribution is given by
\begin{equation}
\mathscr{F}_s^{(4)} = \frac{T}{2}\sum_{\textbf k \alpha}\frac{|\phi_s + \phi_d|^4}{\left(E_{\alpha}(\vec k)^2 + k_n^2\right)^2}.
\end{equation}
\begin{figure}
\includegraphics[width=0.8\columnwidth]{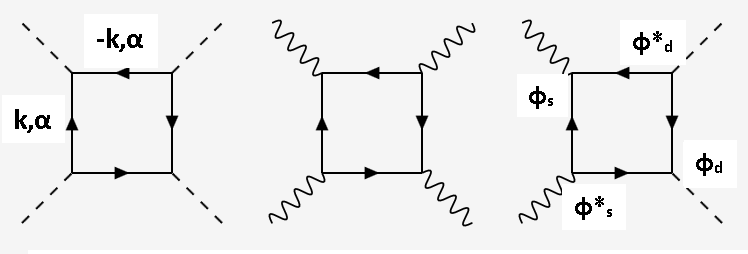}
\caption{Feynman diagrams contributing to the free energy which are independent of the relative phases of the two constituent order parameters. The dashed and wiggly lines represent $d-$wave and $s-$wave cooper pairing. }
\end{figure}
Note that here $\phi_s$ and $\phi_d$ have the respective momentum dependences from the point group of the crystal. Thus, we define $\Delta_s$ and $\Delta_d$ by $\phi_{s,d} \equiv \Delta_{s,d} \Phi_{s,d}(\vec k)$ with the functions $\Phi_{s,d}(\vec k)$ containing all the momentum dependence. We now evaluate these contributions to the free energy by expanding into the quadratic and quartic powers of $\Delta_s$ and $\Delta_d$ to give
\begin{eqnarray}\nonumber
\mathscr{F}_s &=& \alpha_s | \Delta_s |^2 + \alpha_d | \Delta_d |^2 + \beta_s | \Delta_s |^4 +\beta_d | \Delta_d |^4 + 4 \beta_1 | \Delta_s |^2 | \Delta_d |^2  \\ \nonumber
&&+ \alpha' (\Delta_s^* \Delta_d + \Delta_s \Delta_d^*) + \beta_2 \left(\Delta_s^2 \Delta_d^{* 2} + \Delta_s^{*2} \Delta_d^{2} \right) \\
&& + 2 \sum_{\nu=\pm}g_{\nu} (| \Delta_s |^2 + \nu | \Delta_d |^2)  ( \Delta_s \Delta_d^* + 2 \Delta_s^* \Delta_d),
\label{FreeEnergy}
\end{eqnarray} 
with the following definitions of the coefficients (with $g_{\pm} \equiv (g_s \pm g_d)/2$)
\begin{eqnarray}
\alpha_{s,d} &=& T \sum_{\textbf k \alpha} \frac{\Phi_{s,d}^2(\vec k)}{ E_{\alpha}(\vec k)^2 + k_n^2}\\
\alpha' &=& T \sum_{\textbf k \alpha} \frac{\Phi_{s}(\vec k)\Phi_{d} (\vec k)}{ E_{\alpha}(\vec k)^2 + k_n^2}\\
\beta_{s,d} &=& T \sum_{\textbf k \alpha} \frac{\Phi_{s,d}^4(\vec k)}{ \left(E_{\alpha}(\vec k)^2 + k_n^2\right)^2}\\
\beta_1 =\beta_2 &=&  T \sum_{\textbf k \alpha} \frac{\Phi_s(\vec k)^2 \Phi_d(\vec k)^2}{\left(E_{\alpha}(\vec k)^2 + k_n^2 \right)^2}\\
g_{s,d} &=& T\sum_{\textbf k \alpha} \frac{\Phi_{s,d}^3(\vec k)\Phi_{d,s}(\vec k)}{\left(E_{\alpha}(\vec k)^2 + k_n^2 \right)^2}.
\end{eqnarray}
\begin{figure}
\includegraphics[width=0.8\columnwidth]{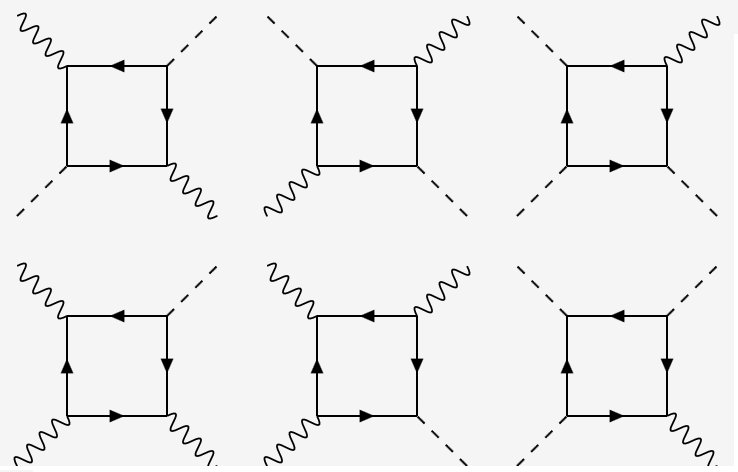}
\caption{Feynman diagrams contributing to the free energy which depend on the relative phases of the two constituent order parameters. The dashed and wiggly lines represent $d-$wave and $s-$wave cooper pairing. }
\end{figure}
Given that $\alpha'$ and $g_{\pm}$ are small compared to the other coefficients due to negligible overlap between odd powers of $\Phi_s(\vec k)$ and $\Phi_d(\vec k)$, and for simplicity in minimizing the free energy, it is easier to study the problem in the absence of linear couplings of $\Delta_s$ and $\Delta_d$, i.e we ignore terms in the free energy proportional to $ (\Delta_s^* \Delta_d + \Delta_s \Delta_d^*)$.   Defining a phase difference between the $d-$wave and $s-$wave gaps to be equal to $\eta$, the free energy is minimized when $\eta = \pi/2$. Minimization of $\mathscr{F}_s$ with respect to $| \Delta_s|$ and $| \Delta_d |$ yields two sets of equations for $\Delta_s$ and $\Delta_d$ given by (defining $\alpha_{s,d} = \alpha\pm \delta \alpha$)
\begin{eqnarray*}
| \Delta_s | \left(\alpha+\delta \alpha + 2 \beta_s | \Delta_s | ^2 + 2 \beta_1 | \Delta_d |^2\right)&=&0\\
| \Delta_d | \left(\alpha-\delta \alpha + 2 \beta_d | \Delta_d | ^2 + 2 \beta_1 | \Delta_s |^2\right)&=&0.
\end{eqnarray*}
Solving these equations gives us four combinations of solutions $\left(|\Delta_s| , |\Delta_d|\right) = (0,0)$,                 $\left(|\Delta_s| , |\Delta_d|\right) = (0,\sqrt{\frac{-(\alpha - \delta \alpha)}{2 \beta_d}})$, $\left(|\Delta_s|, |\Delta_d|\right) = (\sqrt{\frac{-(\alpha + \delta \alpha)}{2 \beta_s}},0)$ and $\left(|\Delta_s|, |\Delta_d|\right) =\left( \sqrt{\frac{\alpha(\beta_1 - \beta_d) - \delta \alpha (\beta_1 + \beta_d)}{2 (\beta_s\beta_d - \beta_1^2)}},\sqrt{\frac{\alpha(\beta_s - \beta_1) - \delta \alpha(\beta_s + \beta_1)}{2(\beta_1^2 -\beta_s\beta_d)}}\right)$. The last pair of solutions which gives rise to a time reversal symmetry broken (TRSB) phase ($|\Delta_s| + i | \Delta_d |$) has the lowest free energy but is obviously a stable solution only when $(| \Delta_s |, | \Delta_d |)$ are both real. Hence, the conditions $\frac{\alpha(\beta_1 - \beta_d) - \delta \alpha (\beta_1 + \beta_d)}{2 (\beta_s\beta_d - \beta_1^2)}>0$ and $\frac{\alpha(\beta_s - \beta_1) - \delta \alpha(\beta_s + \beta_1)}{2(\beta_1^2 -\beta_s\beta_d)}>0$ define the TRSB phase. The $s-$wave and $d-$wave pairings are stable when $-(\alpha+\delta \alpha)$ and $-(\alpha-\delta \alpha)$ are positive respectively. These phases are shown in  Fig. 3 (panels a-c) of the main text. The presence of the linear terms can be treated perturbatively and yields Fig 3 (d) as described in the main text when $\beta_s = \beta_d$.
\end{document}